\documentclass[10pt, conference, letterpaper]{IEEEtran}
\IEEEoverridecommandlockouts
\usepackage{cite}
\usepackage{algorithm}
\usepackage{algorithmic}
\usepackage{graphicx}
\usepackage{textcomp}
\usepackage{xcolor}
\def\BibTeX{{\rm B\kern-.05em{\sc i\kern-.025em b}\kern-.08em
    T\kern-.1667em\lower.7ex\hbox{E}\kern-.125emX}}

\usepackage{amsmath,amssymb,amsfonts, amsthm}
\usepackage{bm}
\usepackage{ulem}
\usepackage{algorithmic}
\graphicspath{{figs/}}
\newtheorem{assumption}{Assumption}
\newtheorem{theorem}{Theorem}
\newtheorem{corollary}{Corollary}

\newtheorem{definition}{Definition}

\DeclareMathOperator{\E}{\mathbb{E}}
\DeclareMathOperator*{\argmin}{argmin}
\DeclareMathOperator*{\argmax}{argmax}

\newcommand{\normsq}[1]{\left\lVert#1\right\rVert^2}
\newcommand{\abs}[1]{\lvert#1\rvert}
\newcommand{\absq}[1]{\left\lvert#1\right\rvert^2}
\usepackage{mathtools}

\DeclarePairedDelimiterX{\innp}[2]{\langle}{\rangle}{#1,#2}

\def\MSE{\mbox{\textrm{MSE}}}
\allowdisplaybreaks

\begin{document}
\title{Federated Learning  over Wireless Networks: \\ A Band-limited Coordinated Descent Approach
{}
}

\author{\IEEEauthorblockN{Junshan Zhang\textsuperscript{1}, Na Li\textsuperscript{2}, and Mehmet Dedeoglu\textsuperscript{3}}
\IEEEauthorblockA{\textsuperscript{1,3}School of Electrical, Computer and Energy Engineering, Arizona State University\\
\textsuperscript{2}School of Engineering and Applied Sciences, Harvard University\\
\textsuperscript{1}Junshan.Zhang@asu.edu, \textsuperscript{2}nali@seas.harvard.edu, \textsuperscript{3}Mehmet.Dedeoglu@asu.edu}
}


\maketitle

\begin{abstract}
We consider a many-to-one wireless architecture for federated learning at the network edge, where multiple edge devices collaboratively train a  model using local data. The unreliable nature of wireless connectivity, together with constraints in computing resources at edge devices, dictates that the local updates at edge devices should be carefully crafted and compressed to match the wireless communication resources available and should work in concert with the receiver. Thus motivated, we propose SGD-based bandlimited coordinate descent algorithms for such settings. Specifically, for the wireless edge employing over-the-air computing,  a common subset of k-coordinates of the gradient updates across edge devices are  selected by the receiver in each iteration, and  then transmitted simultaneously over k sub-carriers, each experiencing time-varying channel conditions.  We characterize the impact of communication error and compression, in terms of the resulting gradient bias and mean squared error, on the convergence of the proposed algorithms. We then study learning-driven communication error minimization via joint optimization of power allocation and learning rates. Our findings reveal that optimal power allocation across different sub-carriers should take into account both the gradient values and channel conditions, thus generalizing the widely used water-filling policy. We also develop sub-optimal distributed solutions amenable to implementation.

\end{abstract}

\section{Introduction}
\setlength\abovedisplayskip{1pt}
\setlength\belowdisplayskip{1pt}
In many edge networks, mobile and IoT devices collecting a huge amount of data are often connected to each other or a central node wirelessly. The unreliable nature of wireless connectivity, together with constraints in computing resources at edge devices,  puts forth a significant challenge for the computation, communication and coordination required to learn an accurate  model at the network edge.  In this paper, we consider a many-to-one wireless architecture for distributed learning at the network edge, where the edge devices collaboratively train a machine learning model, using local data, in a distributed manner. This departs from conventional  approaches which rely heavily on cloud computing to handle high complexity processing tasks, where one  significant  challenge is to meet the stringent low latency requirement. Further, due to privacy concerns, it is highly desirable to derive local learning model updates without sending data to the cloud. In such distributed learning scenarios, the communication between the edge devices and the server can become a bottleneck, in addition to the other challenges in achieving edge intelligence. 

In this paper, we consider a wireless edge network with $M$ devices  and an edge server, 
where a high-dimensional machine learning model is trained using distributed learning.  In such a setting with unreliable and rate-limited communications, local updates at sender devices should be carefully crafted and compressed to make full use of the wireless communication resources available and should work in concert with the receiver (edge server) so as to learn an accurate model.  
Notably, lossy wireless communications for edge intelligence presents unique challenges and opportunities \cite{Zhu2018a}, subject to bandwidth and power requirements, on top of the employed multiple access techniques. Since it often suffices to compute a function of the sum of the local updates for  training  the  model, over-the-air computing is a favorable alternative to the standard multiple-access communications for edge learning. More specifically, over-the-air computation \cite{Goldenbaum2013, Abari2016} takes advantage of the superposition property of wireless multiple-access channel via simultaneous analog transmissions of the local messages, and then computes a function of the messages at the receiver, scaling signal-to-noise ratio (SNR) well with an increasing number of users. In a nutshell, when multiple edge devices collaboratively train a model, it is plausible to employ distributed learning over-the-air. 


We seek to answer the following key questions: 1) What is the impact of the wireless communication bandwidth/power on the accuracy and convergence of the edge learning? 
2) What coordinates in local gradient signals  should be communicated by each edge device to the receiver? 
3) How should the coordination be carried out so that multiple sender devices can work in concert with the receiver? 4) What is the optimal way for the receiver to process the received noisy gradient signals to be used for the stochastic gradient descent algorithm?   5) How should each sender device carry out power allocation across subcarriers to transmit its local updates? Intuitively, it is sensible to allocate more power to a coordinate with larger gradient value to speed up the convergence. Further, power allocation should also be channel-aware.

\begin{figure*}[!tbh]
	\begin{center}
		\centerline{\includegraphics[width=1.5\columnwidth]{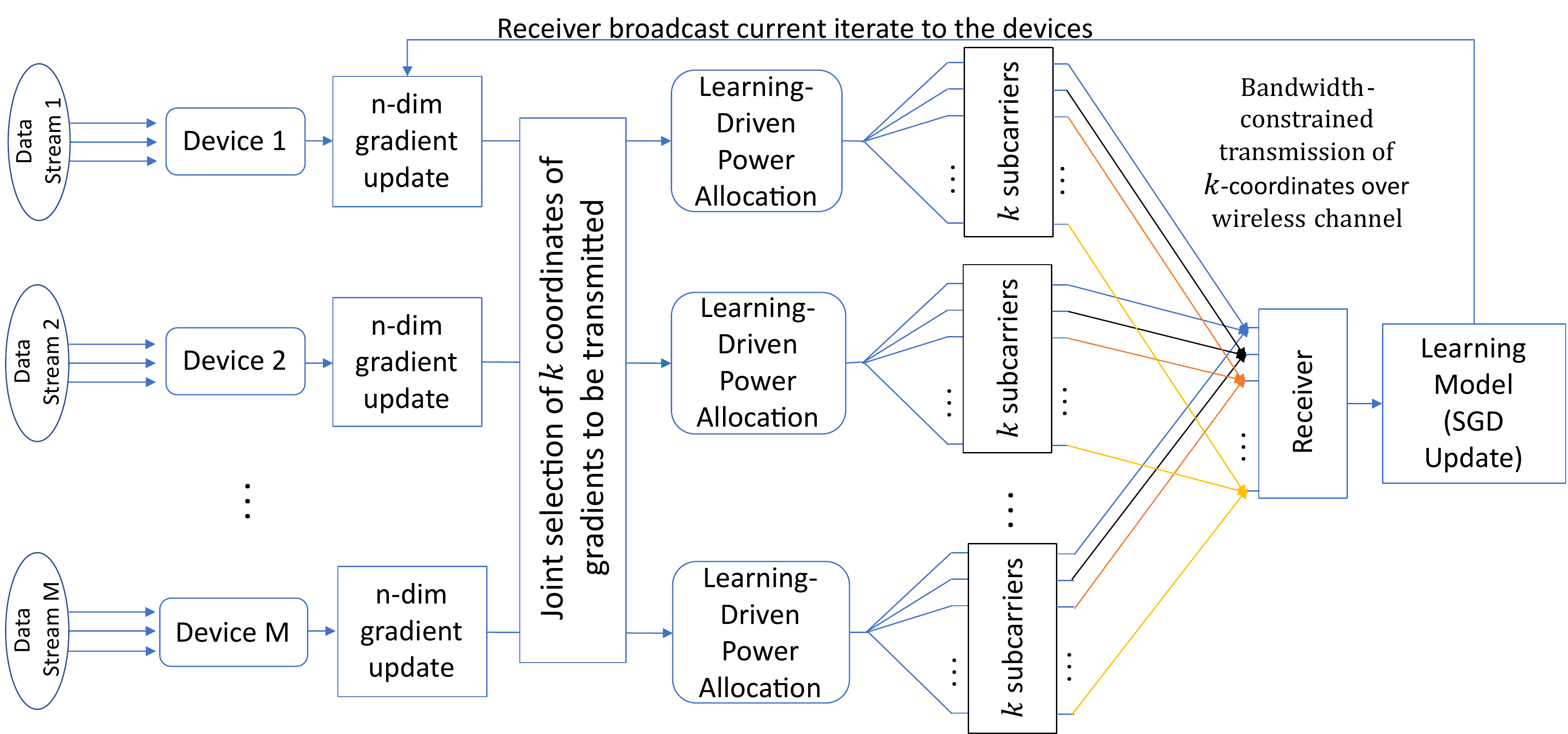}}
		\caption{A bandlimited coordinate descent algorithm for distributed learning over wireless multi-access channel}
		\label{commsmodel}
	\end{center}
	\vspace{-7ex}
\end{figure*}

To answer the above questions, we  consider an integrated learning and communication scheme where multiple edge devices send their local gradient updates over multi-carrier communications to the receiver for learning. Let $K$ denote the number of subcarriers for communications, where $K$ is determined by the wireless bandwidth. First, $K$ dimensions of the gradient updates are determined (by the receiver) to be transmitted.    Multiple methods can be used for selecting $K$ coordinates, e.g., selecting the top-$k$ (in absolute value) coordinates of the sum of the gradients or randomized uniform selection.  This paper will focus on randomly uniform selection (we elaborate further on this in Section V).  During the subsequent communications, the gradient updates are transmitted only in the $K$-selected dimensions via over-the-air computing over $K$ corresponding sub-carriers, each experiencing time-varying channel conditions and hence time-varying transmission errors. The devices are subject to power constraints, giving rise to a key question on how to allocate transmission power across dimension, at each edge device,  based on the gradient update values and channel conditions. Thus, we explore joint optimization  of the power allocation and the learning rate to obtain the best estimate of the gradient updates and minimize the impact of the communication error. We investigate a centralized solution to this problem as a benchmark, and then devise sub-optimal distributed solutions amenable to practical implementation. We note that we have also studied the impact of errors of synchronization across devices in this setting (we omit the details due to limited space).


The main contributions of this paper are summarized as follows:
\begin{itemize}
    
\item  We take a holistic approach to study federated learning algorithms over wireless MAC channels, and the proposed bandlimited coordinated descent(BLCD)  algorithm is built on innovative integration of computing in the air, multi-carrier communications, and wireless resource allocation.

 \item We characterize the impact of communication error and compression, in terms of its resulting gradient bias and mean squared error (MSE), on the convergence performance of the proposed algorithms. Specifically,
when the communication error is unbiased,  the BLCD algorithm would converge to a stationary point under very mild conditions on the loss function.  
In the case the bias in the communication error does exist,
the iterates of the BLCD algorithm would return to a contraction region centered around a scaled version of the bias infinitely often.
 
 \item To minimize the impact of the communication error, we  study joint optimization of power allocation at individual devices and learning rates at the receiver. Observe that since there exists tradeoffs between bias and variance,  minimizing the MSE of the communication error does not necessarily amount to minimizing the bias therein.  Our findings reveal that optimal power allocation across different sub-carriers should take into account both the gradient values and channel conditions, thus generalizing the widely used water-filling policy. We also develop sub-optimal distributed solutions amenable to implementation.
 In particular, due to the power constraints at individual devices, it is not always feasible to achieve  unbiased estimators of the gradient signal across the coordinates. To address this complication, we develop a distributed algorithm 
 which can drive the bias  in the communication error to (close to) zero under given power constraints and then  reduce the corresponding variance as much as possible. 
 
\end{itemize}

\section{Related Work}

 Communication-efficient SGD algorithms are of great interest to reduce latency caused by the transmission of the high dimensional gradient updates with minimal performance loss. Such algorithms in the ML literature are based on compression via quantization \cite{Alistarh2016, Wen2017, Bernstein2018a, Wu2018}, sparsification \cite{Aji2017, Stich2018, Alistarh2018} and federated learning \cite{Konecny2016} (or local updates \cite{Stich2018a}), where lossless communication is assumed to be provided. At the wireless edge, physical-layer design and communication loss should be taken into consideration for the adoption of the communication-efficient algorithms.

Power allocation for over-the-air computation is investigated for different scenarios in many other works \cite{Dong2018, Liu2019, Wen2018, Zhu2018b, Cao2019} including MIMO, reduced dimensional MIMO, standard many to one channel and different channel models. In related works on ML over wireless channels, \cite{Zhu2018, Yang2019, Zeng2019, Amiri2019, Amiri2019a, Amiri2019c, Ahn2019, Sery2019} consider over-the-air transmissions for training of the ML model. The authors in \cite{Amiri2019} propose sparsification of the updates with compressive sensing for further bandwidth reduction, and recovered sum of the compressed sparse gradients is used for the update. They also apply a similar framework for federated learning and fading channels in \cite{Amiri2019a}. \cite{Zhu2018} considers a broadband aggregation for federated learning with opportunistic scheduling based on the channel coefficients for a set of devices uniformly distributed over a ring. Lastly, \cite{Sery2019} optimize the gradient descent based learning over multiple access fading channels. It is worth noting that the existing approaches for distributed learning in wireless networks do not fully account for the characteristics of lossy wireless channels. It is our hope that the proposed BLCD algorithms can lead to an innovative architecture of distributed edge learning over wireless networks that accounts for computation, power, spectrum constraints and packet losses.

\section{Federated Learning over Wireless Multi-access Networks}

\subsection{Distributed Edge Learning Model}
Consider an edge computing environment with $M$ devices $\mathcal{M}=\{1,\ldots,M\}$ and an edge server. As illustrated in Figure 1, a high-dimensional ML model is trained at the server by using an SGD based algorithm, where stochastic gradients are calculated at the devices with the data points obtained by the devices and a (common) subset of the gradient updates are transmitted through different subcarriers via over-the-air.	

The general edge learning problem is as follows:
\begin{equation}
	\min_{w\in\mathbb{R}^d} f(w):=\frac{1}{M} \sum_{m=1}^{M} \mathbb{E}_{\xi_m} [ l(w, \xi_m)],
\end{equation}
in which $l(\cdot)$ is the loss function, and edge device $m$ has access to inputs  $\xi_m $.
Such optimization is typically performed through empirical risk minimization iteratively. In the sequel, we let $w_t$ denote the parameter value of the ML model at communication round $t$, and at round $t$ edge device $m$ uses its local data $\xi_{m,t}$ to  compute a stochastic gradient $g^m_t (w_t):=\nabla l(w_t,\xi_{m,t})$. Define $g_t(w_t) = \frac{1}{M}\sum_{m=1}^{M} g^m_t (w_t)$.
The standard vanilla SGD algorithms is given as
\begin{equation} \label{eqn:genericupdate}
w_{t+1} = w_t - \gamma g_t(w_t)
\end{equation}
with $\gamma$ being the learning rate.  Nevertheless, different updates can be employed for different SGD algorithms, and this study will focus on communication-error-aware SGD algorithms.

\subsection{Bandlimited Coordinate Descent Algorithm}

Due to the significant discrepancy between the wireless bandwidth constraint and the high-dimensional nature of the gradient signals, we propose a sparse variant of the SGD algorithm over wireless multiple-access channel, named as bandlimited coordinate descent (BLCD), in which at each iteration  only a common set of $K$ coordinates, $I(t)\subset \{1, \ldots, d\}$ (with $K\ll d$), of the gradients are selected to be transmitted through over-the-air computing for the gradient updates. The details of coordinate selection for the BLCD algorithm are relegated to Section \ref{sec:controlphase}. Worth noting is that due to the unreliable nature of wireless connectivity,  the communication is assumed to be lossy, resulting in erroneous estimation of the updates at the receiver. Moreover, gradient correction is performed by keeping the difference between the update made at the receiver and the gradient value at the transmitter for the subsequent rounds, as gradient correction dramatically improves the convergence rate with sparse gradient updates \cite{Stich2018}. 

For convenience, we first define the gradient sparsification operator as follows.
\begin{definition}
	$ C_I : \mathbb{R}^d \rightarrow \mathbb{R}^d$ for a set $I \subseteq \{1,\ldots, d\}$ as follows: for every input $x \in \mathbb{R}^d$, $	\big(C_I (x)\big)_j$ is $(x)_{j} $ for $ j \in I$ and $0$ otherwise.
\end{definition}
Since this operator $C_I$ compress a $d$-dimensional vector to a $k$-dimension one, we will also refer this operator as compression operator in the rest of the paper. 

\begin{algorithm}[!t]
	\caption{Bandlimited Coordinate Descent Algorithm}\label{alg_1}
	\begin{algorithmic}[1]
		\STATE \textbf{Input:} Sample batches \(\xi_{m,t}\), model parameters \(w_1\), initial learning rate \(\gamma\), sparsification operator \(C_t(.)\), \(\forall m=1,\dots,M; \forall t=1,\dots,T.\)
		\STATE \textbf{Initialize:} \(r_t^m:=0\).
		\FOR{$t=1:T$}
    		\FOR{$m=1:M$} 
    		\STATE \(g_t^m(w_t):= \text{stochasticGradient}(f(w_t,\xi_{m,t}))\)
    		\STATE \(u_{t}^m := \gamma g_t^m(w_t)+r_t^m \)
    		\STATE \(r_{t+1}^m := u_t^m-C_t(u_t^m)\)
    		\STATE Compute power allocation coefficients \(b_{km}^*,\forall k=1,\dots,K\).
    		\STATE Transmit \(\mathbf{b}^*\odot C_t(u_t^m)\)
    		\ENDFOR
		\STATE Compute gradient estimator $\hat{G}_t(w_t)$
		\STATE \(w_{t+1}:= w_t -  \hat{G}_t(w_t)  \).
		\STATE Broadcast \(w_{t+1}\) back to all transmitters.
		\ENDFOR
	\end{algorithmic}
\end{algorithm}

With a bit abuse of notation, we let $C_t$ denote $C_{I(t)}$  for convenience  in the following.  Following \cite{Karimireddy2019}, we  incorporate the sparsification error made in each iteration (by the compression
operator $C_t$) into the next step to alleviate the possible gradient bias therein and  improve the convergence possible. Specifically, as in \cite{Karimireddy2019},  one plausible way for compression error correction is to 
update the gradient correction  term as follows:
\begin{align} 
    r_{t+1}^m &= u_t^m - C_t(u_t^m), \label{eqn:SGDmemupdatestd}\\
   u_t^m &\triangleq \gamma g^m_t(w_t) + r_t^m 
\end{align}
which $ r_{t+1}^m $ keeps the error in the  sparsification operator that is in the memory of user $m$ at around $t$, and  $u_t^m $ is the scaled gradient with correction at device $m$ where the scaling factor $\gamma$ is the learning rate in equation~\eqref{eqn:genericupdate}. {(We refer readers to \cite{Karimireddy2019} for more insights of this error-feedback based compression SGD.)}
 Due to the lossy nature of wireless communications, 
there would be communication errors and  the gradient estimators at the receiver would be erroneous. In particular, the gradient estimator at the receiver in the  BLCD can be written as
\begin{equation} \label{eqn:SGDupdate}
\hat{G}_t(w_t) = \frac{1}{M}\sum_{m=1}^{M} C_t \left(u_t^m \right) + \epsilon_t,
\end{equation}
where $\epsilon_t$ denotes the random communication error in round $t$. 
In a nutshell, the bandlimited coordinate descent algorithm
is outlined in Algorithm~\ref{alg_1}.

Recall that $g_t(w_t) = \frac{1}{M}\sum_{m=1}^{M} g^m_t(w_t)$ and define $r_t \triangleq \frac{1}{M}\sum_{m=1}^{M} r^m_t$.
Thanks to the common sparsification operator across devices,   the update in the SGD algorithm at communicatioon round $t$ is given by 
\begin{equation} \label{eqn:updatesimplified}
w_{t+1} = w_t -  \big[ C_t(\gamma g_t(w_t) +r_t) + \epsilon_t\big]. 
\end{equation}
To quantify the impact of the communication error, we use the corresponding communication-error free counterpart as the benchmark, defined as follows:
\begin{equation} \label{eqn:gcsimplified}
 \hat{w}_{t+1} = w_t -   C_t(\gamma g_t(w_t) +r_t) . 
\end{equation}
It is clear that $w_{t+1}=  \hat{w}_{t+1} - \epsilon_t $.

For convenience, we define \(\tilde{w}_t \triangleq {w}_{t} - r_t \).
It can be shown that 
\(\tilde{w}_{t+1}=  \tilde{w}_{t} -  \gamma g_t(w_t) - \epsilon_{t} \).
Intuitively, $w_{t+1}$  in (\ref{eqn:updatesimplified}) is a noisy version of  the  iterate $\hat{w}_{t+1}$ in (\ref{eqn:gcsimplified}), which implies that  
\(\tilde{w}_{t+1} \) is a noisy version of the compression-error correction  of $\hat{w}_{t+1}$ in (\ref{eqn:gcsimplified}), where the ``noisy perturbation'' is incurred by the communication error. 

\subsection{BLCD Coordinate Transmissions over Multi-Access Channel}

\begin{figure}[h]
	\begin{center}
		\centerline{\includegraphics[width=\columnwidth]{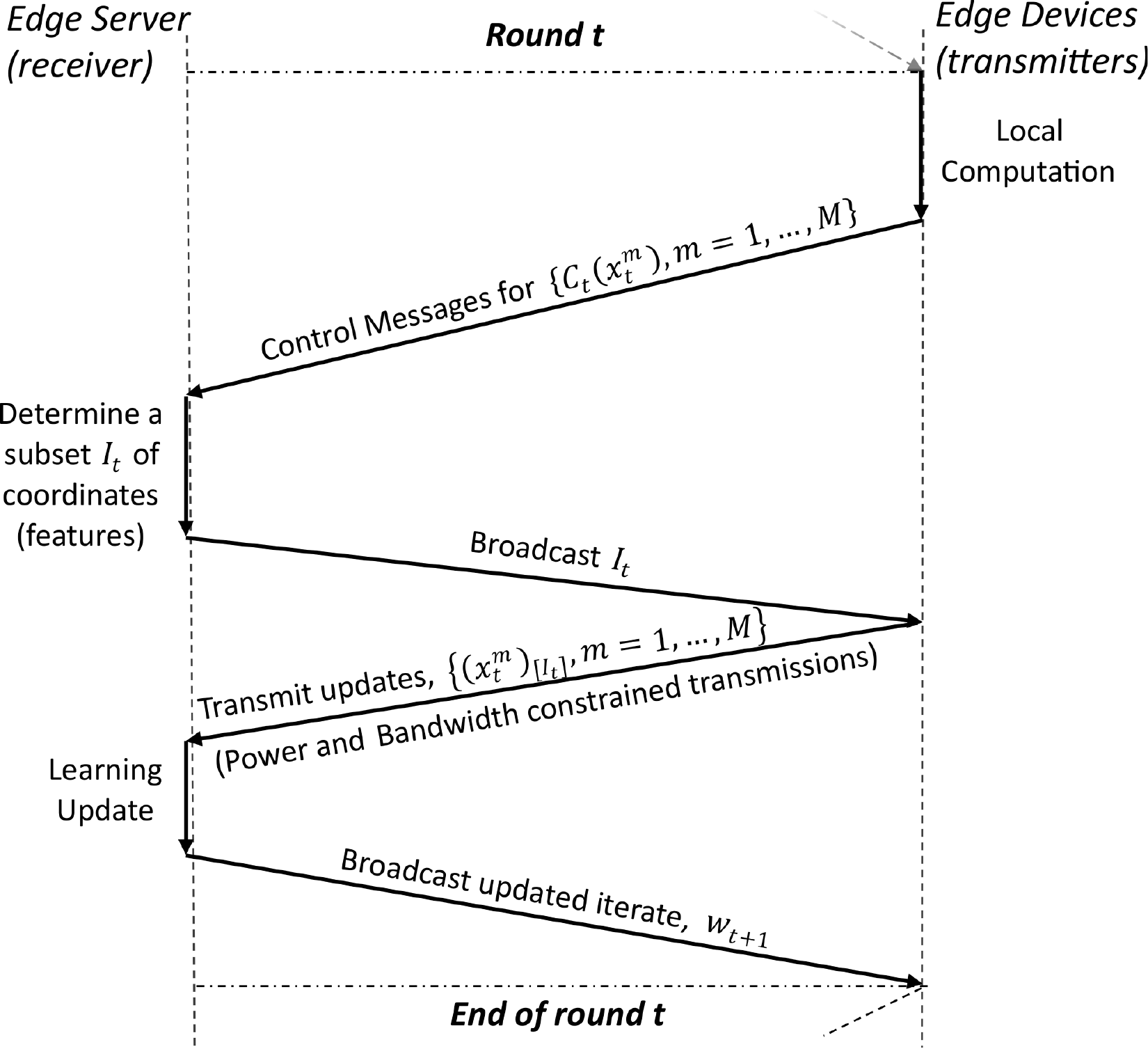}}%
		\vspace{0in}
        \caption{A multi-access communication protocol for bandlimited coordinate selection and transmission.}\label{flowchart1}
	\end{center}
	\vspace{-0.15in}
\end{figure}


A key step in the BLCD algorithm is to achieve coordinate
synchronization of the transmissions among many edge devices. 
To this end,  we introduce a receiver-driven low-complexity   multi-access communication protocol, as illustrated in Fig.~\ref{flowchart1}, with the function $C_t(x)$ denoting the compression of $x$ at round $t$. Let $I(t)$ (of size $K$) denote the subset of coordinates  chosen for transmission  by the receiver at round $t$.   Observe  that the updates at the receiver are carried out only in the dimensions $I(t)$. Further,  the edge
receiver can broadcast its updated iterate to
participant devices, over the reverse link. 
This task is quite simple, 
given the broadcast nature of wireless channels. 
In the transmissions, each coordinate of the gradient updates is mapped to a specific subcarrier and then transmitted through the wireless MAC channel, and the coordinates transmitted by different devices over the same subcarrier are received by the edge server in the form of an aggregate sum. {It is worth noting that the above protocol is also applicable to the case when the SGD updates are carried out for multiple rounds at the devices.} 


When there are many edge devices, over-the-air computation 
can be used to take advantage of superposition property of wireless multiple-access channel 
via simultaneous analog transmissions of the local updates. More specifically, 
at round t, the received signal in subcarrier $k$ is given by:
\begin{equation} \label{eqn:channel}
y_k (t) = \sum_{m=1}^{M} b_{km}(t) h_{km}(t)  x_{km} (t) + n_k (t)
\end{equation}
where $b_{km}(t)$ is a power scaling factor, $h_{km}(t)$ is the channel gain, and $x_{km}(t)$ is the message of user $m$ through the subcarrier $k$, respectively, and $n_k (t) \sim \mathcal{N}(0,\sigma^2)$ is the channel noise. 

To simplify notation, we omit $(t)$ when it is clear from the context in the following. Specifically, the message $x_{km}=(C_t(u^m_t))_{l(k)}$, with a  one-to-one mapping $l(k)=(I(t))_k$, which indicates the $k$-th element of $I(t)$, transmitted through the $k$-th subcarrier.
The total power that a device can use in the transmission is limited in practical systems. Without loss of generality, we assume that there is a power constraint at each device, given by $\sum_{k=1}^{K} \absq{b_{km} x_{km}} \leq E_m,\ \forall m\in \{ 1, \ldots, M \}$.
Note that $b_{km}$ hinges heavily upon both $\bm{h}_m=[h_{1m}, \ldots, h_{Km}]^\top$ and $\bm{x}_{m}=[x_{1m}, \ldots, x_{Km}]^\top$, and a key next step is to optimize $b_{km} (\bm{h}_{m}, \bm{x}_{m})$. In each round, each device optimizes its power allocation for transmitting the selected coordinates of its update signal over the $K$ subcarriers, aiming to minimize the communication error so as to achieve a good estimation of $G_t(w_t)$ (or its scaled version) for  the gradient update, where 
$$G_t(w_t) \triangleq \frac{1}{M}\sum_{m=1}^M C_t(u_t^m).$$ 


From the learning perspective, based on $\{y_k\}_{k=1}^K$, it is of paramount importance for the receiver to get a good estimate of $G_t(w_t)$. Since $n_k(t)$ is Gaussian noise, the optimal estimator is in the form of \vspace{0.05in}
\begin{equation} \label{eqn:estimator}
\big(\widehat{G}_t(w_t)\big)_{k} = 
\begin{cases} 
\alpha_{l(k)} y_{l(k)}, & k \in I(t)  \\
0 & \text{otherwise}
\end{cases}\vspace{0.05in}
\end{equation} 
where $\{ \alpha_k \}_{k=1}^K$ are gradient estimator coefficients for subcarriers. It follows that the communication error (i.e., the  gradient estimation error incurred by lossy communications) is given by
\begin{equation}
\epsilon_t = \widehat{G}_t(w_t) - G_t(w_t) . \label{comm-error}
\end{equation}
We note that  $\{\alpha_k\}_{k=1}^K$  are intimately related to the learning rates for the $K$ coordinates, scaling the learning rate to be  $\{\gamma \alpha_k\}_{k=1}^K$. It is interesting to observe that  the learning rates in the proposed BLCD algorithm are essentially different across the dimensions, due to the unreliable and dynamically changing channel conditions across different subcarriers.

\section{Impact of  Communication Error and Compression on BLCD Algorithm}	\label{sec:convergence}

Recall that due to the common sparsification operator across devices,  the update in the SGD algorithm at communication round $t$ is given by 
\[ 
w_{t+1} = w_t -  \big[ C_t(\gamma g_t(w_t) +r_t) + \epsilon_t\big]. 
\]
Needless to say, the compression operator $C_t$ plays a critical role in sparse transmissions. In this study, we impose the following standard assumption on the compression rate of the operator. 
\begin{assumption} \label{asmpt:compression}
	For a set of the random compression operators $\{C_t\}_{t=1}^T$ and any $x\in \mathbb{R}^d$, it holds
	\begin{equation}
	\E \normsq{x - C_t(x)} \leq (1-\delta) \normsq{x}
	\end{equation}
	for some $\delta \in (0,1]$.
\end{assumption}
  
We impose the following standard assumptions on the  non-convex objective function $f(\cdot)$ and the corresponding stochastic gradients $g^m_t (w_t)$ computed with the data samples of device $m$ in round $t$. (We assume that the data samples $\{\xi_{m,t}\}$
are i.i.d.~across the devices and time.)
\begin{assumption} \label{asmpt:smoothness}
	(Smoothness) A function $f:\mathbb{R}^d \rightarrow \mathbb{R}$ is L-smooth if for all ${x},{y}\in \mathbb{R}^d$, it holds
	\begin{equation}
	\abs{f({y})-f({x})-\innp{\nabla f({x})}{{y}-{x}}} \leq \frac{L}{2} \normsq{{y}-{x}}.
	\end{equation}
\end{assumption}
\begin{assumption} \label{asmpt:boundedmoment}
	For any $x\in \mathbb{R}^d$ and for any $m=1, \ldots, M$, a stochastic gradient $g_t^m(x), \forall t$, satisfies
	    \begin{equation}
	        \E[g_t^m(x)] = \nabla f (x), \textrm{ } \E \normsq{g_t^m(x)} \leq G^2
	    \end{equation}
	where $G>0$ is a constant.
\end{assumption}

\begin{table*}[t]
	\centering
	\begin{minipage}{1\textwidth}
		\begin{align}
   \mathbb{E}_t [ f(\tilde{w}_{t+1}) ] 
   \hspace{-0.03in}\leq& f(\tilde{w}_{t} )\hspace{-0.03in}+\hspace{-0.03in}\langle\nabla f(\tilde{w}_{t}),\mathbb{E}_t[\tilde{w}_{t+1}\hspace{-0.03in}-\hspace{-0.03in}\tilde{w}_{t}]\rangle  \hspace{-0.03in}+\hspace{-0.03in}\frac{L}{2}\mathbb{E}_t[\lVert \tilde{w}_{t+1}-\tilde{w}_{t} \rVert^2] \nonumber\\
   &\hspace{-0.5in}=f(\tilde{w}_{t})\hspace{-0.03in}-\hspace{-0.03in} \langle\nabla f(\tilde{w}_{t}),\gamma \mathbb{E}_t[g_t (w_t)]  \hspace{-0.03in}+\hspace{-0.03in} \mathbb{E}_t[\epsilon_t] \rangle \hspace{-0.03in}+\hspace{-0.03in} \frac{L}{2}\mathbb{E}_t[\lVert\gamma g_t(w_t) \rVert^2]\hspace{-0.03in}+\hspace{-0.03in}\frac{L}{2}\mathbb{E}_t[\lVert\epsilon_t\rVert^2] \hspace{-0.03in}+\hspace{-0.03in} L\mathbb{E}_t[\langle \gamma g_t(w_t) ,\epsilon_t\rangle]\nonumber\\
    &\hspace{-0.5in}= f(\tilde{w}_{t}) \hspace{-0.03in}-\hspace{-0.03in}\langle\nabla f({w}_{t}),\gamma \mathbb{E}_t[g_t(w_t)] \hspace{-0.03in}+\hspace{-0.03in} \mathbb{E}_t[\epsilon_t]\rangle  \hspace{-0.03in}-\hspace{-0.03in} \langle\nabla f(\tilde{w}_{t})\hspace{-0.03in}-\hspace{-0.03in}\nabla f({w}_{t}),  \gamma \mathbb{E}_t[g_t(w_t)] \hspace{-0.03in}+\hspace{-0.03in} \mathbb{E}_t[\epsilon_t]\rangle \hspace{-0.03in}+\hspace{-0.03in} \frac{L}{2}\mathbb{E}_t[\Vert\epsilon_t\rVert_2^2] \hspace{-0.03in}+\hspace{-0.03in} L\mathbb{E}_t[\langle\gamma g_t(w_t),\epsilon_t\rangle] \hspace{-0.03in}+\hspace{-0.03in} \frac{L}{2}\mathbb{E}_t[\Vert\gamma g_t(w_t) \rVert^2 \nonumber\\
    &\hspace{-0.5in}\leq f(\tilde{w}_{t}) \hspace{-0.03in}-\hspace{-0.03in} \gamma \lVert \nabla f(w_t) \rVert_2^2 \hspace{-0.03in}-\hspace{-0.03in} \langle \nabla f(w_t), \mathbb{E}_t[\epsilon_t] \rangle \hspace{-0.03in}+\hspace{-0.03in} \frac{\rho}{2} \lVert \gamma \nabla f(w_t)\hspace{-0.03in}+\hspace{-0.03in}\mathbb{E}_t[\epsilon_t] \rVert_2^2\hspace{-0.03in}+\hspace{-0.03in}\frac{L^2}{2\rho}\mathbb{E}_t[\lVert r_t \rVert_2^2] \hspace{-0.03in}+\hspace{-0.03in} \frac{L}{2}\mathbb{E}_t[\lVert \epsilon_t \rVert_2^2] \hspace{-0.03in}+\hspace{-0.03in} L\langle \nabla f(w_t), \mathbb{E}_t[\epsilon_t] \rangle \hspace{-0.03in}+\hspace{-0.03in} \frac{L\gamma^2}{2} \mathbb{E}_t \lVert g_t(w_t) \rVert_2^2 \nonumber\\
    &\hspace{-0.5in}\leq f(\tilde{w}_{t}) \hspace{-0.03in}-\hspace{-0.03in} \gamma \lVert \nabla f(w_t) \rVert_2^2 \hspace{-0.03in}+\hspace{-0.03in} (L-1) \lVert \nabla f(w_t) \rVert\lVert \mathbb{E}_t[\epsilon_t] \rVert \hspace{-0.03in}+\hspace{-0.03in} \frac{\rho}{2}\left( \gamma^2\lVert \nabla f(w_t) \rVert_2^2 \hspace{-0.03in}+\hspace{-0.03in} \lVert \mathbb{E}_t[\epsilon_t] \rVert_2^2\hspace{-0.03in}+\hspace{-0.03in}2\gamma \langle \nabla f(w_t), \mathbb{E}_t[\epsilon_t] \rangle \right) \hspace{-0.03in}+\hspace{-0.03in} \frac{L^2}{2\rho} \mathbb{E}_t[\lVert r_t \rVert_2^2] \hspace{-0.03in}+\hspace{-0.03in} \frac{L}{2}\mathbb{E}_t[\lVert\epsilon_t\rVert_2^2] \hspace{-0.03in}+\hspace{-0.03in} \frac{L\gamma^2}{2} G^2 \nonumber\\
    &\hspace{-0.5in}\leq f(\tilde{w}_{t}) - \gamma \lVert \nabla f(w_t) \rVert_2^2 + (L-1+ 2 \gamma) \lVert \nabla f(w_t) \rVert\lVert \mathbb{E}_t[\epsilon_t] \rVert + \frac{\gamma^2 \rho }{2}\lVert \nabla f(w_t) \rVert_2^2 + \frac{L^2}{2\rho} \mathbb{E}_t[\lVert r_t \rVert_2^2] + \lVert \mathbb{E}_t[\epsilon_t] \rVert_2^2  +\frac{L}{2}\mathbb{E}_t[\lVert\epsilon_t\rVert_2^2] + \frac{L\gamma^2}{2} G^2 \nonumber\\
    &\hspace{-0.5in}= f(\tilde{w}_{t}) -\gamma \left[ 1-\frac{\rho}{2}\gamma  \right] \lVert \nabla f(w_t) \rVert_2^2 + (L-1+2\gamma) \lVert \nabla f(w_t) \rVert \lVert \mathbb{E}_t[\epsilon_t] \rVert + \frac{L^2}{2\rho} \mathbb{E}_t[\lVert r_t \rVert_2^2] + \lVert \mathbb{E}_t[\epsilon_t] \rVert_2^2  +\frac{L}{2}\mathbb{E}_t[\lVert\epsilon_t\rVert_2^2] + \frac{L\gamma^2}{2} G^2 \label{gradient-bound}
\end{align}
		\hrule
	\end{minipage}\vspace{-0.2in}
\end{table*}

It follows directly from \cite{Karimireddy2019} that
$
    \mathbb{E} [\lVert r_t \rVert_2^2]\leq \frac{4(1-\delta)}{\delta^2}\gamma^2 G^2.
$
Recall that
\(\tilde{w}_{t+1}=  \tilde{w}_{t} -  \gamma g_t(w_t) - \epsilon_{t} \)
and that \(\tilde{w}_{t+1} \) can be viewed as a noisy version of the compression-error correction  of $\hat{w}_{t+1}$ in (\ref{eqn:gcsimplified}), where the ``noisy perturbation'' is incurred by the communication error.   For convenience, let $ \mathbb{E}_t[\epsilon_t]$ denote    the gradient bias incurred by the communication error    and $ \mathbb{E}_t [\lVert \epsilon_t \rVert_2^2 ]$ be  the corresponding mean square error, where $ \mathbb{E}_t$ is taken with respect to channel noise.

Let $\eta= \frac{L-1+2\gamma}{\gamma (2-\rho\gamma)} $ with  $0<\rho<2$. Let $f^*$ denote the globally minimum value of $f$.
We have the following main result on the iterates in the BLCD algorithm.

\begin{theorem} \label{thm:convergence}
	Under Assumptions \ref{asmpt:compression}, \ref{asmpt:smoothness} and \ref{asmpt:boundedmoment}, the iterates $\{w_t\}$ in the BLCD algorithm  satisfies that
\begin{align}
&\frac{1}{T\hspace{-0.03in}+\hspace{-0.03in}1}\sum_{t=0}^T \left(\lVert\nabla f(w_t)\rVert_2 \hspace{-0.03in}-\hspace{-0.03in}  \eta \lVert \underbrace{\mathbb{E}_t[\epsilon_t]}_{\mbox{bias}} \rVert_2\right)^2
    \nonumber\\
    &\hspace{0.1in}\leq\hspace{-0.03in}\frac{1}{T\hspace{-0.03in}+\hspace{-0.03in}1}\sum_{t=0}^T \left[\frac{L\eta }{ L \hspace{-0.03in}-\hspace{-0.03in}1 \hspace{-0.03in}+\hspace{-0.03in} 2 \gamma}\underbrace{\mathbb{E}_t[\lVert\epsilon_t\rVert_2^2]}_{\mbox{MSE}} \hspace{-0.03in}+\hspace{-0.03in} \left(1\hspace{-0.03in}+\hspace{-0.03in} \eta^2 \right)\hspace{-0.03in} \lVert \underbrace{\mathbb{E}_t[\epsilon_t]}_{\mbox{bias}} \rVert_2^2 \right]\nonumber\\
    &\hspace{0.1in}+\hspace{-0.03in}\frac{2}{T\hspace{-0.03in}+\hspace{-0.03in}1}\frac{f(w_0)\hspace{-0.03in}-\hspace{-0.03in}f^*}{\gamma(2\hspace{-0.03in}-\hspace{-0.03in}\rho\gamma)}\hspace{-0.03in}+\hspace{-0.03in} \left(\frac{L}{\rho} \frac{2(1\hspace{-0.03in}-\hspace{-0.03in} \delta)}{\delta^2} \hspace{-0.03in}+\hspace{-0.03in} \frac{1}{2} \right)\hspace{-0.03in} \frac{2L \gamma G^2}{ 2-\rho\gamma}.  \label{main-result}
\end{align}
\end{theorem}
\vspace{-0ex}

\begin{proof}
Due to the limited space, we outline only a few main steps for the proof. Recall that \(\tilde{w}_t= {w}_{t} - r_t \).
It can be shown that 
\(\tilde{w}_{t+1}=  \tilde{w}_{t} -  \gamma g_t(w_t) - \epsilon_{t} \). As shown in (\ref{gradient-bound}),
using the properties of the iterates in the BLCD algorithm and the smoothness of the objective function $f$,
 we can establish an upper bound on
$\mathbb{E}_t [ f(\tilde{w}_{t+1}) ] $
in terms of \(f(\tilde{w}_{t}) \) the corresponding gradient \( \nabla f(w_t)  \), and the gradient bias and MSE due to the communication error. Then, (\ref{main-result}) can be obtained after some further algebraic manipulation.
\end{proof}

 {\bf Remarks.} Based on  Theorem~\ref{thm:convergence}, we have a few observations in order. 
 \begin{itemize}
 
  \item We first examine the four terms on the right hand side of (\ref{main-result}): The first two terms  capture the impact on the gradient by the time average of the bias in the communication error $\epsilon_t$ and that of the corresponding the mean square, denoted as MSE; the two items would go to zero if the bias and the MSE diminish; the third term is a scaled version of $ f(w_0) - f^* $ and would go to zero as long as $\gamma  = O(T^{-\beta}) $ with $\beta < 1$; and the fourth term is proportional to $\gamma$ and would go to zero when $\gamma \rightarrow 0$. 

\item If the right hand side of (\ref{main-result}) diminishes as $T \rightarrow \infty$, 
 the iterates in the BLCD algorithm would ``converge'' to a neighborhood around  $\eta \lVert \mathbb{E}_t[\epsilon_t] \rVert_2$, which is a scaled version of the bias in the communication error.  For convenience, let $ \bar{\epsilon}= \limsup_t  \lVert \mathbb{E}_t[\epsilon_t] \rVert_2$, and
 define a contraction region as follows:
\[
A_{\gamma} = \left\{ w_t:  \lVert\nabla f(w_t)\rVert_2 \leq  (\eta + \Delta) \bar{\epsilon} \right\}.
\]
where $\Delta >0$ is an arbitrarily  small positive number.  It then follows that the iterates in the BLCD algorithm would ``converge'' to a contraction region given by $A_{\gamma}$, in the sense that the iterates return to $A_{\gamma}$ infinitely often. 
Note that $f$ is assumed to be any nonconvex smooth function, and there can be many contraction regions, each corresponding to a stationary point.

\item When the communication error is unbiased, the gradients  would diminish to $0$ and hence the BLCD algorithm would converge to a stationary point.  
In the case the bias in the communication error does exist,
there exists intrinsic tradeoff between the size of the contraction region and $\eta \lVert \mathbb{E}_t[\epsilon_t] \rVert_2$. When the learning rate $\gamma$ is small, the   right hand side of (\ref{main-result}) would small, but $\eta$ can be large, and vice verse.
It makes sense to choose a fixed learning rate that would make $\eta$ small. In this way, the gradients in the BLCD algorithm would  ``concentrate" around  a (small) scaled version of the bias.  

\item  Finally,  the impact of gradient sparsification is captured by $\delta$. For instance, when (randomly) uniform selection is used,   $\delta=\frac{k}{d}$. We will elaborate on this 
 in Section \ref{sec:controlphase}. 
 
 \end{itemize}
 
 Further, we have the following corollary. 
\begin{corollary} \label{thm:convergencezeromean}
	Under Assumptions \ref{asmpt:compression}, \ref{asmpt:smoothness}, and \ref{asmpt:boundedmoment}, we have that
if $\E_t[\epsilon_t]=0$ and \(\gamma = \frac{1}{\sqrt{T+1}} \), the BLCD algorithm converges to a stationary point and satisfies that
 \begin{align}
    &\frac{1}{T\hspace{-0.03in}+\hspace{-0.03in}1}\hspace{-0.03in}\sum_{t=0}^T \lVert\nabla f(w_t)\rVert_2^2 \nonumber\\
    &\leq\hspace{-0.03in} \frac{1} {2 - \frac{\rho}{\sqrt{T+1}} }
    \left\{ \frac{2  (f(w_0)\hspace{-0.03in}-\hspace{-0.03in}f^*) }{\sqrt{T+1}} \hspace{-0.03in}+\hspace{-0.03in}  \frac{2 L G^2}{\sqrt{T\hspace{-0.03in}+\hspace{-0.03in}1}} \hspace{-0.03in}\left(\frac{L}{\rho} \frac{2(1\hspace{-0.03in}-\hspace{-0.03in}\delta)}{\delta^2} \hspace{-0.03in}+\hspace{-0.03in}  \frac{1}{2} \right)\hspace{-0.03in} \right. \nonumber\\
    &\left. \hspace{0.5in}+\ \frac{L}{T+1}
    \sum_{t=0}^T\underbrace{\mathbb{E}_t[\lVert\epsilon_t\rVert_2^2]}_{\text{MSE}} \right\}
\end{align}  	 
\end{corollary}
\vspace{-2ex}


\section{Communication Error Minimization via Joint Optimization of Power Allocation and Learning Rates} 
\label{sec:update}	


Theorem~\ref{thm:convergence}
reveals that the communication error  has a significant impact on the convergence behavior of the BLCD algorithm. In this section, we turn our attention to minimizing the  communication error (in term of MSE and bias) via joint optimization of power allocation and learning rates.

Without loss of generality, we focus on iteration $t$ (with abuse of notation, we omit $t$ in the notation for simplicity). 
Recall that the coordinate updates in the BLCD algorithm, sent by different devices over the same subcarrier,  are received by the edge server as an aggregate sum, which is used to estimate the gradient value in that specific dimension. We denote the power coefficients and estimators as $\bm{b} \triangleq [b_{11}, b_{12}, \ldots, b_{1M}, b_{21}, \ldots, b_{KM} ]$ and $\bm{\alpha} \triangleq  [\vec{\alpha}_{1}, \ldots, \vec{\alpha}_{K}]$. In each round, each sender device optimizes its power allocation for transmitting the selected coordinates of their updates over the $K$ subcarriers, aiming to achieve the best convergence rate. 
We assume that the perfect channel state information is available at the corresponding transmitter, i.e.,  $\bm{h}_m=[h_{1m}, \ldots, h_{Km}]^\top$ is available at the sender $m$ only.

Based on (\ref{comm-error}), the mean squared error of the communication error in iteration $t$  is given by
\begin{equation}
 \mathbb{E}_t [\lVert \epsilon_t \rVert_2^2 ]  = \E\bigg[\normsq{\widehat{G}_t(w_t) - G_t(w_t)}\bigg]
\end{equation}
where the expectation is taken over the channel noise.
For convenience, we denote $\mathbb{E}_t [\lVert \epsilon_t \rVert_2^2 ] $ as $\MSE_1$, and after some algebra, it  can be rewritten as the sum of the variance and the square of the bias: 
\begin{align}
\hspace{-0.08in} \MSE_1(\bm{\alpha}, \bm{b})\hspace{-0.03in}=\hspace{-0.05in} \sum_{k=1}^K\hspace{-0.04in}  \bigg[\hspace{-0.04in} \underbrace{\sum_{m=1}^M\hspace{-0.04in}  \left(\hspace{-0.03in}  \alpha_k b_{km} h_{km} \hspace{-0.04in} -\hspace{-0.04in} \frac{1}{M}\hspace{-0.04in}  \right)\hspace{-0.04in}  x_{km}\hspace{-0.04in} }_{\mbox{bias in $k$th coordinate}} \bigg]^2\hspace{-0.11in}  +\hspace{-0.04in}   \underbrace{\sum_{k=1}^K \hspace{-0.04in} \sigma^2  \alpha^2_k}_{\mbox{variance\ }   }
\label{bias-var}
\end{align}
Recall that  $\{\alpha_k\}_{k=1}^K$  are intimately related to the learning rates for the $K$ coordinates, making the learning rate effectively  $\{\gamma \alpha_k\}_{k=1}^K$.

\subsection{Centralized Solutions to Minimizing MSE (Scheme 1)}

In light of the above,  we can cast the MSE minimization problem as a learning-driven joint power allocation and learning rate problem, given by
\begin{align}
\textbf{P1:\ } \min_{\bm{\alpha}, \bm{b}} \quad & \MSE_1(\bm{\alpha}, \bm{b}) \\
\textrm{s.t.} \quad & \sum_{k=1}^{K} \absq{ b_{km} x_{km} } \leq E_m, \quad \forall m \\
& b_{km}\geq 0, \ \alpha_k \geq 0 \quad \forall k,m
\end{align}
which minimizes the MSE for every round. 

The above formulated problem is non-convex because the objective function involves the product of variables. Nevertheless, it is biconvex, i.e., for one of the variables being fixed, the problem is convex for the other one. 
In general, we can solve the above bi-convex optimization problem in the same spirit as in the EM algorithm, by taking the following two steps, each optimizing over a single variable, iteratively:
\begin{equation*}
\begin{aligned}
 \textbf{P1-a:} \ \min_{\bm{\alpha}} \quad & \MSE_1(\bm{\alpha}, \bm{b})\quad \textrm{s.t.} \quad \alpha_k \geq 0, \quad \forall k \\
 \textbf{P1-b:} \min_{\bm{b}} \quad & \MSE_{11}(\bm{\alpha}, \bm{b}) \\
\textrm{s.t.} \quad & \sum_{k=1}^{K} \absq{ b_{km} x_{km} } \leq E_m \ \ \forall m, \quad b_{km}\geq 0 \ \ \forall k,m.
\end{aligned} 
\end{equation*} 	
Since (\textbf{P1-a}) is unconstrained,for given 
$\{ b_{km} \} $, 
the optimal solution to (\textbf{P1-a}) is given by
	\begin{align}
	\alpha^*_k \! = \! \max \left\{ \frac{\big(\sum_{m=1}^M x_{km}\big)\big(\sum_{m=1}^M b_{km} h_{km} x_{km} \big)}{M \big[\sigma^2 + \big(\sum_{m=1}^M b_{km} h_{km} x_{km} \big)^2\big]}, 0\right\}.
	\end{align}	
Then, we can solve (\textbf{P1-b}) by  optimizing $\bm{b}$ only. Solving the sub-problems (\textbf{P1-a}) and (\textbf{P1-b}) iteratively leads to a local minimum, however, not necessarily to the global solution.

Observe that the above solution requires the global knowledge of $x_{km}$'s and $h_{km}$'s of all devices, which is difficult to implement in practice. We will treat it as a {\em  benchmark} only. Next, we  turn our attention to developing distributed sub-optimal solutions.

\vspace{-0.05in}
\subsection{Distributed Solutions towards Zero Bias and Variance Reduction (Scheme 2)}

As noted above, the centralized solution to ({\bf P1}) requires the global knowledge of $x_{km}$'s and $h_{km}$'s and hence is not amenable to implementation. Further, minimizing the MSE of the communication error does not necessarily amount to minimizing the bias therein since  there exists tradeoffs between bias and variance. Thus motivated,  we next focus on devising distributed sub-optimal solutions which can drive the bias  in the communication error to (close to) zero, and then  reduce the corresponding variance as much as possible. 

Specifically, observe from (\ref{bias-var}) that the minimization of MSE cost does not necessarily ensure \(\hat{G}\) to be an unbiased estimator, due to the intrinsic tradeoff between bias and variance. To this end, we take a sub-optimal approach where the optimization problem is decomposed into two subproblems. In the subproblem at the transmitters, each device  \(m\) utilizes its available power and local gradient/channel information to compute a power allocation policy in terms of \(\{b_{1m},b_{2m},\dots,b_{Km}\}\). In the subproblem at the receiver, the receiver  finds the best possible \(\alpha_k\) for all \(k=1,\dots,K\). Another complication is that due to the power constraints at individual devices, it is not always feasible to achieve  unbiased estimators of the gradient signal across the coordinates. Nevertheless, for given power constraints, one can achieved   unbiased estimators of a scaled down version of the coordinates  of the gradient signal.  In light of this, we formulate the optimization problem at each device (transmitter) $m$ to ensure an unbiased estimator of a scaled  version $\zeta_m$ of the transmitted  coordinates,   as follows:
\begin{align}\label{prob2a}
 \mbox{\bf Device~m:} \  &\underset{\{b_{km}\}_{k=1:K}}{\max} ~ \zeta_m\\
   \text{s.t.}\hspace{0.1in}\sum_{k=1}^K b_{km}^2&x_{km}^2\leq E_m, \ \  b_{km}\hspace{-0.03in}\geq\hspace{-0.03in} 0, \\
   \zeta_mx_{km}&-b_{km}h_{km}x_{km}= 0, & \forall k=1,\dots,K,\label{q56}
\end{align}
 where maximizing  $\zeta_m$  amounts to maximizing the corresponding SNR (and hence improving the gradient estimation accuracy).
The first constraint in the above is the power constraint, and the second constraint is imposed to ensure that there is no  bias of the same scaled   version of the transmitted signals across the dimensions for user $m$.  The power allocation solution can be found using Karush-Kuhn-Tucker (KKT) conditions as follows:
\begin{align}\label{q57}
\Aboxed{\zeta_m^* = \sqrt{\frac{E_m}{\sum_{k=1}^K \frac{x_{km}^2}{h_{km}^2}}},~ b_{km}^*=\frac{\zeta^*_{m}}{h_{km}}, ~\forall k.}
\end{align}

Observe  that  using the obtained power allocation policy in (\ref{q57}), all $K$ transmitted coordinates for device $m$  have the same scaling factor $\zeta_m$. Next, we will ensure zero bias by choosing the right \(\boldsymbol{\alpha}\) for gradient estimation at the receiver, which can be obtained by solving the following optimization problem since all transmitted gradient signals are superimposed via the over-the-air transmission:  
\begin{align}
 \mbox{\bf Receiver side:} \ 
    \underset{\{\alpha_k\},}{\min}~ &\sum_{k=1}^K {\nu}_k^2(\alpha_k, \{b_{km}^*\})\label{q58}\\
    \text{s.t.}~ e_k(\alpha_k, \{b_{km}^*\}) = 0, &~~~\alpha_k\geq 0, \forall k=1,\dots,K, \label{q59}
\end{align}
where \(e_k\) and  \( \nu_k^2 \)  denote the bias and variance components, given as follows:
\begin{align}
    e_k(\alpha_k, \{b_{km}^*\}) 
     &= \alpha_k\left( \sum_{m=1}^M \zeta_m^* x_{km} \right) - \frac{1}{M}\sum_{m=1}^M x_{km},\nonumber\\
    \nu_k^2(\alpha_k, \{b_{km}^*\}) &= \alpha_k^2\sigma^2,
\end{align}
for all \(k = 1,\dots,K\). For given  $\{\zeta_m^*\}$, it is easy to see that
\begin{align}
 \Aboxed{ \alpha_k^* = \frac{\frac{1}{M}\sum_{m=1}^M x_{km}}{\sum_{m=1}^M \zeta_m^*x_{km}}  \simeq \frac{1}{\sum_{m=1}^M \zeta_m^*}, \ \ \forall k.}
\end{align}
We note that in the above, from an implementation point of view, since $\{x_{km} \}$ is not available at the receiver, it is sensible to set $ \alpha_k^\dagger
\simeq \frac{1}{\sum_{m=1}^M \zeta_m^*}$. Further, \( \{\zeta_m^*\} \) is not readily available at the receiver either. Nevertheless, since there is only one parameter $\zeta_m^*$ from each sender $m$, the sum \(\sum_{m=1}^M \zeta_m^*\) can be sent over a control channel to the receiver to compute \(\alpha_k^\dagger\).  It is worth noting that in general the bias exists even if $E_m$ is the same for all senders. 

Next, we take a closer look at the case when the number of subchannels \(K\) is large (which is often the case in practice).  Suppose that   \(\{x_{km}\}\) are i.i.d.~across subchannels and users, and so are  \(\{ h_{km} \}\). We  can then simplify \(\zeta_m^*\) further. For ease of exposition, we denote \(\mathbb{E}[x_{km}^2] = \varphi^2+\bar{x}^2\) and \(\mathbb{E}\left[ \frac{1}{h_{km}^2} \right] = \varpi^2\). When \(K\) is large, for every user \(m\) we have that: 
\begin{align}
&\zeta_m^* = \frac{\sqrt{E_m}}{\sqrt{\sum_{k=1}^K \frac{x_{km}^2}{h_{km}^2}}} \underset{\substack{\text{when $K$}\\\text{is large}}}{\Longrightarrow} \zeta_m^*  \approx \frac{\sqrt{E_m}}{\sqrt{K (\varphi^2+\bar{x}^2) \varpi^2}} 
\end{align}
As a result, the bias and variance for each dimension $k$ could be written as,
\begin{align}
e_k(\alpha_k^*, \{b_{km}^*\}) &=  \hspace{-0.03in}\sum_{m=1}^M\hspace{-0.03in} \left[\frac{ \sqrt{E_m}}{\sum_{m=1}^M \sqrt{E_m}} \hspace{-0.03in}-\hspace{-0.03in} \frac{1}{M}\right] x_{km}, \forall k. \label{eq:ek=0}\\
{\nu}_k^2 & =  \hspace{-0.03in}\frac{K\varpi^2(\varphi^2+\bar{x}^2)}{\left(\sum_{m=1}^M\sqrt{E_m}\right)^2}\sigma^2, \forall k. \label{eq82}
\end{align}
 {\em Observe that when $E_m$ is the same across the senders, the bias term \( \mathbb{E}_t [\epsilon_t] =  \mathbf{0}\) in the above setting according to \eqref{eq:ek=0}. }


\subsection{A User-centric Approach Using Single-User Solution (Scheme 3)}

In this section, we  consider  a suboptimal user-centric approach, which  provides insight on the  power allocation across the subcarriers from a single device perspective. We formulate the single device (say user $m$) problem as
\begin{equation*}
\begin{aligned}
\textbf{P2:\ }&\min_{\{b_{km}\}, \{\alpha_k\}}~ \sum_{k=1}^K \bigg[ \big( \alpha_k b_{km} h_{km} - 1 \big) x_{km}\bigg]^2 +  \sigma^2 \sum_{k=1}^K \alpha^2_k \\
&\textrm{s.t.}~ \sum_{k=1}^{K} \absq{ b_{km} x_{km} } \leq E_m; \ b_{k}\geq 0, \ \alpha_k \geq 0, \forall k.
\end{aligned}
\end{equation*}
\vspace{-2ex}

\begin{theorem} \label{thm:beta}
	The optimal solution $\{b_{km}^*, \alpha_k^*\} $ to (\textbf{P2})  is given by
	\begin{equation}
	\label{eqn:singleuser}
	(b^*_{km})^2 = \bigg[ \sqrt{\frac{ \sigma^2} {\lambda x_{km}^2 h_{km}^2} }  - \frac{\sigma^2}{h_{km}^2 x_{km}^2} \bigg]^+, \forall k,
	\end{equation}
	\begin{align}
	    \alpha^*_k=\frac{b^*_{km} h_{km} x^2_{k} }{ \sigma^2 + (b^*_{km})^2 h^2_{km} x^2_{km}}, ~~~~\forall k,
	\end{align}
	where $\lambda_m$ is a key parameter determining the waterfilling level:
		\begin{equation} 
	\sum_{k=1}^K  \bigg[  \sqrt{\frac{1}{\lambda_m}} \sqrt{\frac{x_{km}^2  \sigma^2} { h_{km}^2} }  - \frac{\sigma^2}{h_{km}^2 } \bigg]^+  = E_m. \end{equation}
\end{theorem}

The proof of
Theorem~\ref{thm:beta} is omitted due to space limitation. Observe that Theorem~\ref{thm:beta} reveals that the larger the gradient value (and the smaller channel gain) in one subcarrier, the higher power the it should be allocated to in general, and that $ \{x_{km}/h_{km)} \}$ can be used to compute the water  level for applying the water filling policy.


Based on the above result, in the multi-user setting,  each device can adopt the above single-user power allocation solution as given in Theorem \ref{thm:beta}. This solution can be applied individually without requiring any coordination between devices.

Next, we take a closer look at the case when the number of subchannels \(K\) is large.
Let $\bar{E}_m$ denote the average power constraint per subcarrier. When \(K\) is large, after some algebra, the optimization problem \textbf{P2} can be further approximated as follows:
\begin{align}
    \textbf{P3: } \underset{{b}_{km}}{\min}& ~\mathbb{E} \left[ \frac{{{x}^2_{km}} \sigma^2}
    {{{b}^2_{km}} {{h}^2_{km}} {{x}^2_{km}} +\sigma^2} \right] \nonumber\\
    \text{s.t. }& ~ \mathbb{E}\left[
    {b}^2_{km} {{x}^2_{km}}\right]\leq \bar{E}_m,~ {b}_{km}\geq 0,
\end{align}
where the expectation is taken with respect to $\{{h}_{km}\}$ and $\{{x}_{km}\}$.


The solution for  \(k=1,\dots,K\) is obtained as follows:\vspace{-0.01in}
\begin{align}
  &  b_{km}^* \hspace{-0.04in}=\hspace{-0.04in} \sqrt{\hspace{-0.04in}\left[\hspace{-0.03in} \frac{\sigma \lvert x_{km}\rvert^{-1}}{h_{km}\sqrt{\lambda_m}}\hspace{-0.03in}-\hspace{-0.03in} \frac{\sigma^2}{x_{km}^2h_{km}^2} \hspace{-0.03in}\right]^+}\\
    & \lambda_m \hspace{-0.03in}<\hspace{-0.03in}\frac{h_{km}^2x_{km}^2}{\sigma_k^2}\hspace{-0.03in}\Rightarrow\hspace{-0.03in} b^*_{km}\hspace{-0.03in}>\hspace{-0.03in}0\label{q82}
\end{align}
We can compute the bias and the variance accordingly.

\section{Coordinate Selection for Bandlimited Coordinate Descent Algorithms} \label{sec:controlphase} 

The selection of which coordinates to operate on is crucial to the performance of sparsified SGD algorithms. It is not hard to see that selecting the top-$k$ (in absolute value) coordinates of the sum of the gradients provides the best performance. 
However, in practice it may not always be feasible to obtain top-$k$ of the sum of the gradients, and in fact there are different solutions  for selecting $k$ dimensions with large absolute values; see e.g., \cite{Amiri2019a, Ivkin2019}.
Note that each device individually   transmitting top-$k$ coordinates of their local gradients is not applicable to the scenario of over-the-air communications considered here. Sequential device-to-device transmissions provides an alternative approach \cite{shi2019distributed}, but these techniques are likely to require more bandwidth with wireless connection.

Another  approach that  is considered is the use of compression and/or sketching for the gradients to be transmitted. For instance, in \cite{Amiri2019a}, a system that updates SGD via decompressing the compressed gradients transmitted through over-the-air communication is examined. To the best of our knowledge, such techniques do not come with rigorous convergence guarantees.
A similar approach is taken in \cite{Ivkin2019}, where the sketched gradients are transmitted through an error-free medium and these are then used to obtain top-$k$ coordinates; the devices next simply transmit the selected coordinates. 
Although such an approach can be taken with over-the-air computing since only the summation of the sketched gradients is necessary; this requires the transmission of $\mathcal{O}(k\log d)$ dimensions. To provide guarantees with such an approach $\mathcal{O}(k\log d + k)$  up-link transmissions are needed. Alternatively, uniformly selected $\mathcal{O}(k\log d + k)$ coordinates can be transmitted with similar bandwidth and energy requirements. For the practical learning models with non-sparse updates, uniform coordinate selection tend to perform better. Moreover, the common $K$ dimensions can be selected uniformly via synchronized pseudo-random number generators without any information transfer. To summarize, uniform selection of the coordinates is more attractive based on the energy, bandwidth and implementation considerations compared to the methods aiming to recover top-$k$ coordinates; indeed, this is the approach we adopt.

\section{Experimental Results}

In this section, we evaluate the accuracy and convergence performance of the BLCD algorithm, when using one of the following three schemes for power allocation and learning rate selection (aiming to minimize the impact of communication error): 1) the bi-convex program based solution (Scheme 1), 2) the distributed solution towards zero bias in Section~\ref{sec:update}. (Scheme 2); 3) the single-user solution (Scheme 3).
We use the communication error free scheme as the baseline to evaluate the performance degradation.
We also consider the naive scheme (Scheme 4) using equal power allocation for all dimensions, i.e.,   $b_{km}=\sqrt{E/ {\sum_{k=1}^K x^2_{km}}}$.
 
\begin{figure}[!t]
	\begin{center}
		\includegraphics[width=\columnwidth]{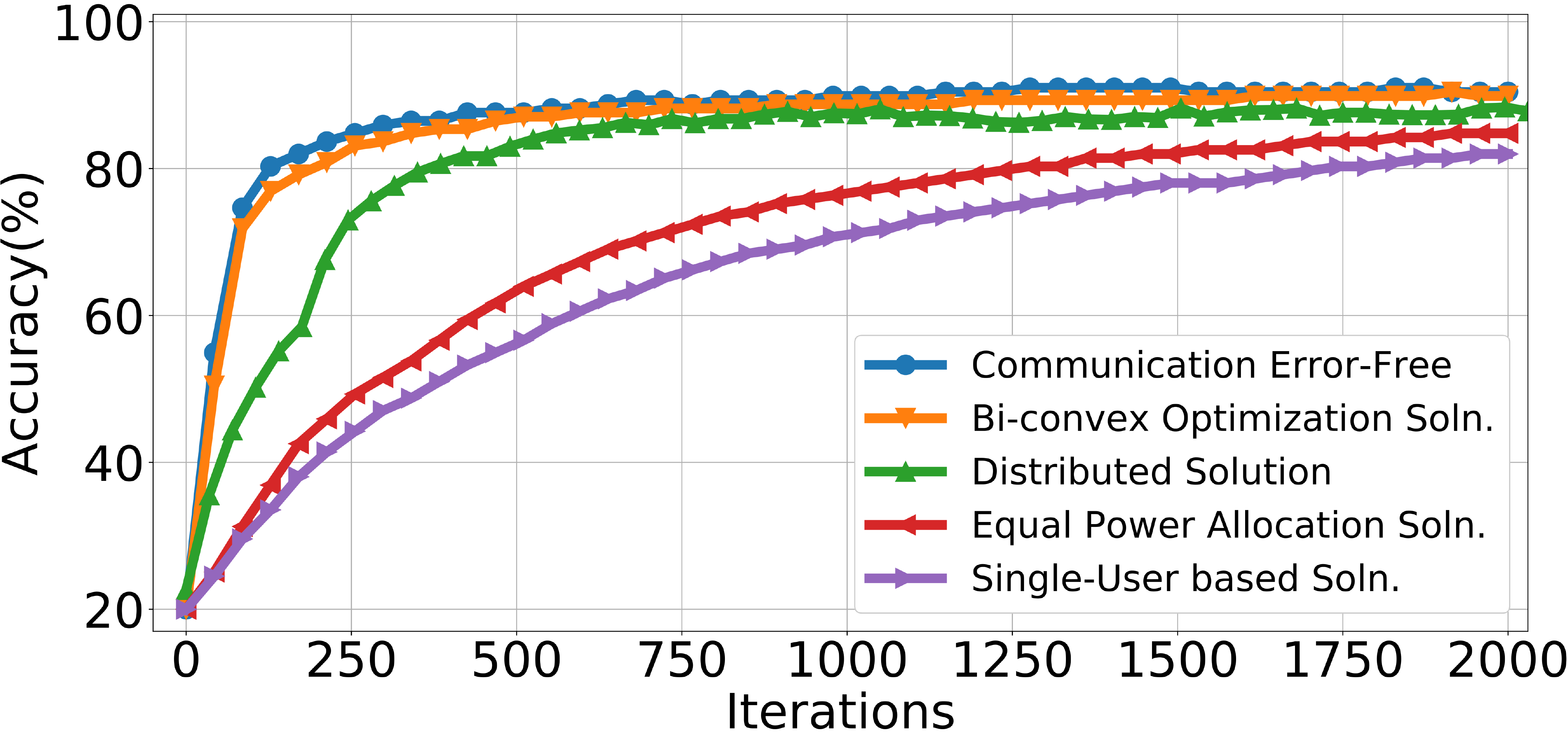} 
		\vspace{-4ex}
		\caption{Testing accuracy over training iterations for $\alpha_k=1/8$, $E_{avg}=0.1$ and a batch size of \(4\). Training model consists of a single layer neural network with 7840 differentiable parameters.} \label{fig:comparison1}
		\vspace{2ex}
		\includegraphics[width=\columnwidth]{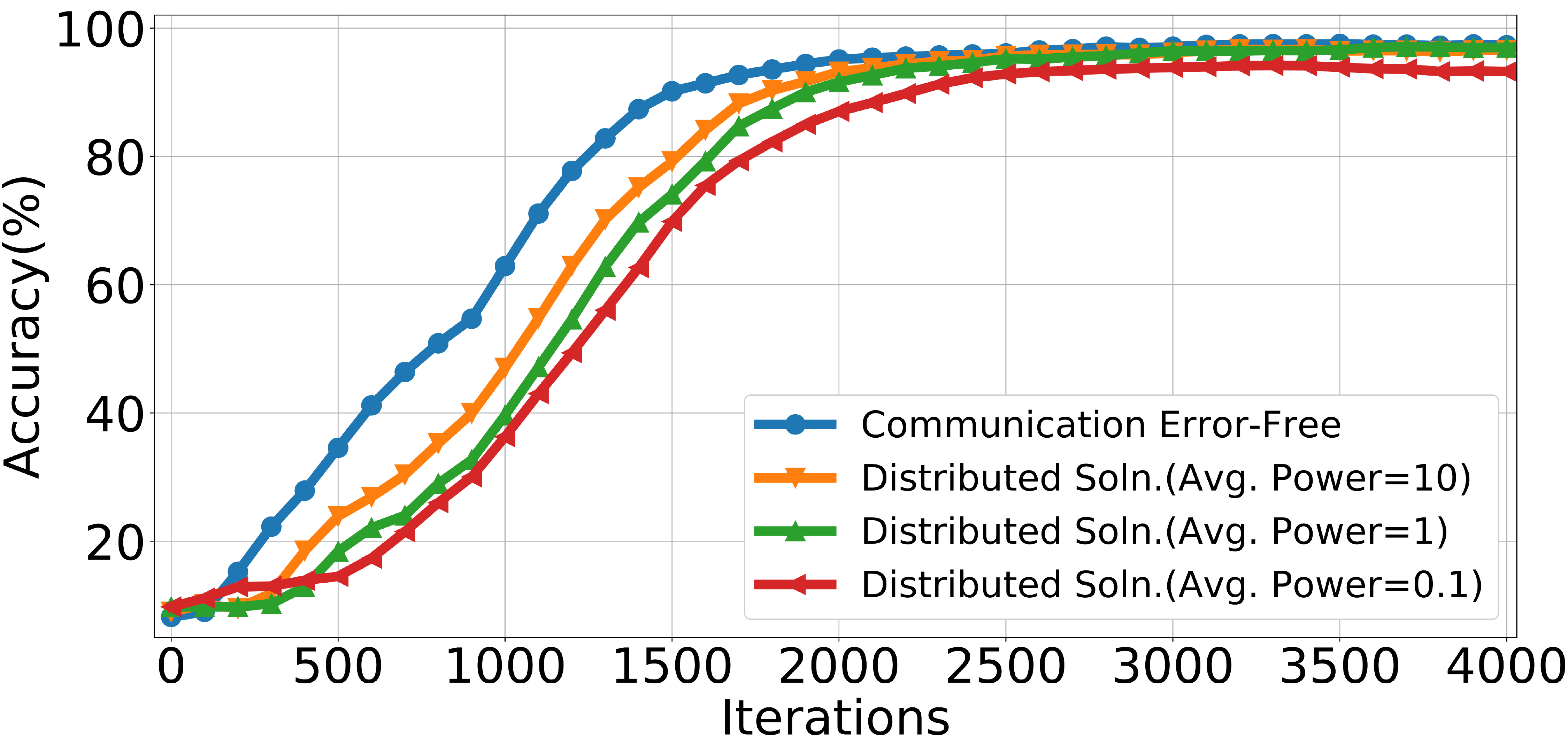} 
		\vspace{-4ex}
		\caption{Testing accuracy over training iterations for $10$ workers and a batch size of \(256\). Training model consists of a \(5\)-layer deep neural network with 61706 differentiable parameters.}\label{fig:comparison2}
		\vspace{-0ex}
	\end{center}
\end{figure}

In our first experiment, we consider a simple single layer neural network trained on the MNIST dataset. The network {consists of two 2-D convolutional layers with filter size \(5\times 5\) followed by a single fully connected layer and it} has 7840 parameters. $K=64$ dimensions are uniformly selected as the support of the sparse gradient transmissions. For convenience, we define $E_{avg}$ as the average sum of the energy (of all devices) per dimension normalized by the channel noise variance, i.e.,
$E_{avg}= E M \E[h_{km}^2]/ K \sigma^2.$ Without loss of generality, we take the variance of the channel noise as $\sigma^2=1$ and $\{h_{km}\}$ are independent and identically distributed Rayleigh random variables with mean $1$. The changes on $E_{avg}$ simply amount to different SNR values. In  Fig. \ref{fig:comparison1}, we take $K=64$, $M=8$, batch size $4$ to calculate each gradient, and the learning rate $\gamma=0.01$. In the second experiment, we expand the model to {a more sophisticated \(5\)-layer neural network and an \(18\)-layer ResNet \cite{resnetpaper} with \(61706\) and \(11175370\) parameters, respectively}. {The \(5\)-layer network consists of two 2-D convolutional layers with filter size \(5\times 5\) followed by three fully connected layers. In all experiments, we have used a learning rate of \(0.01\). the local dataset of each worker is randomly selected  from the entire MNIST dataset.} We use \(10\) workers with varying batch sizes and we utilize \(K=1024\) sub-channels for sparse gradient signal transmission.

It can be seen from Fig.~\ref{fig:comparison1} that in the presence of the communication error, the centralized solution (Scheme 1) based on bi-convex programming converges quickly and performs the best, and it can achieve accuracy close to the ideal error-free scenario. Further, the distributed solution (Scheme 2) can eventually approach the performance of Scheme 1, but the single-user solution (Scheme 3) performs poorly, so does the naive scheme using equal power allocation (Scheme 4). Clearly, there exists significant gap between its resulting accuracy  and that in the error-free case, and this is because the bias in Scheme 3 is more significantly.

\begin{figure}[!t]
	\begin{center}
		\includegraphics[width=\columnwidth]{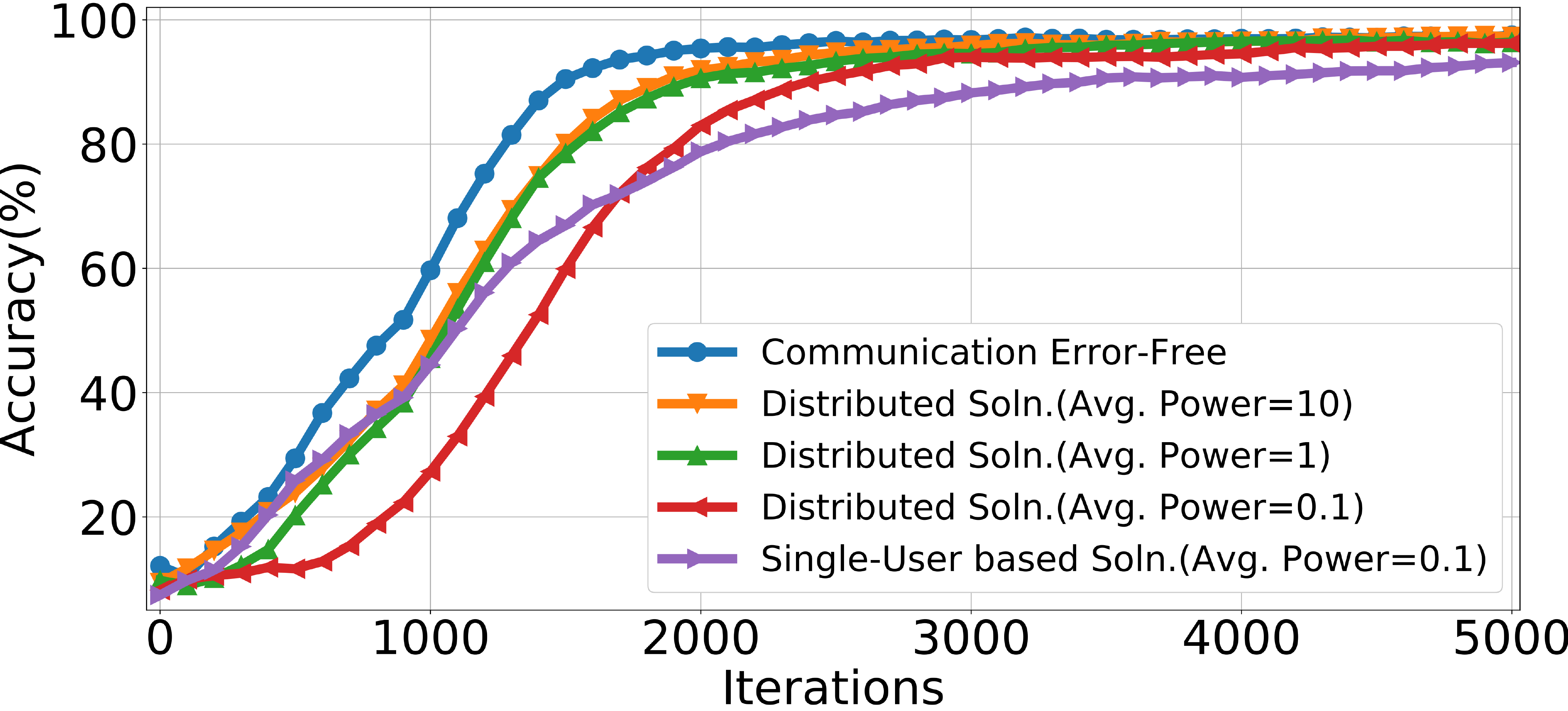} 
		\vspace{-4ex}
		\caption{Testing accuracy over training iterations for $10$ workers and a batch size of \(4\). Training model consists of a \(5\)-layer deep neural network with 61706 differentiable parameters.} \label{fig:comparison3}
		\vspace{2ex}
		\includegraphics[width=\columnwidth]{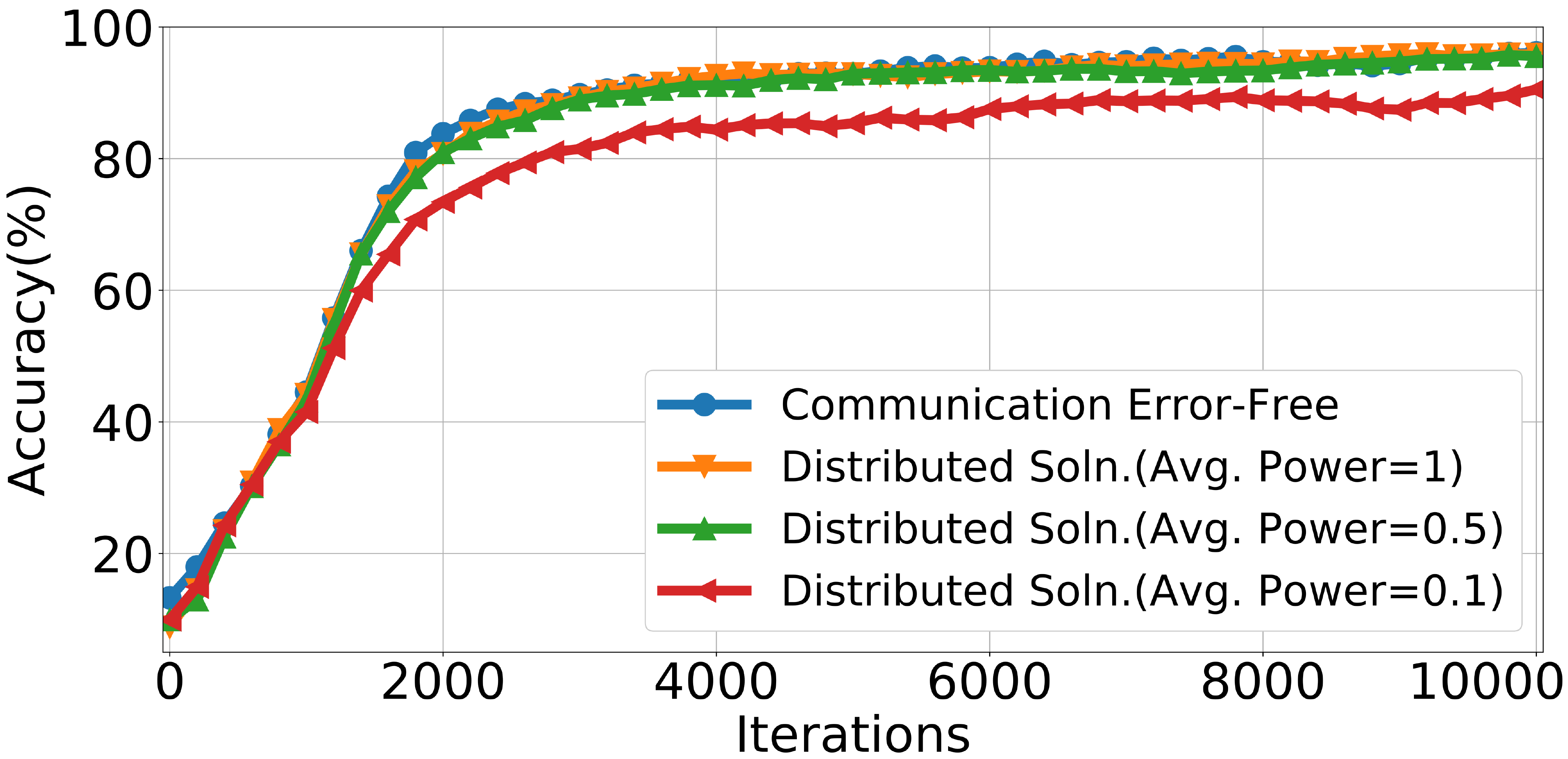} 
		\vspace{-4ex}
		\caption{{Testing accuracy over training iterations for $10$ devices and a batch size of \(4\). Training model consists of an \(18\)-layer ResNet network with more than 11 million differentiable parameters.}}\label{fig:resnet}
	\end{center}\vspace{0ex}
\end{figure}

Next,  Figures~\ref{fig:comparison2}, \ref{fig:comparison3} {and \ref{fig:resnet}} depict the results in the second experiment using much larger-scale deep neural networks. It can be observed from Figs.~\ref{fig:comparison2}, \ref{fig:comparison3} {and \ref{fig:resnet}} that
the SNR can have significant impact on the final accuracy. As expected, {the convergence on the ResNet network is slower in comparison to other DNNs due to the huge number of parameters and small batch size. Nevertheless, it is clear that the learning accuracy improves significantly at high SNR. (The solution of the distributed algorithm for \(E_{avg}=10\) is omitted in Fig. \ref{fig:resnet}, since it is indistinguishably close to error-free solution.}) It is interesting to observe that when the SNR increases, the distributed solution (Scheme 2) can achieve accuracy close to the ideal error-free case, but the single-user solution (Scheme 3) would not. It is worth noting that due to the computational complexity of bi-convex programming in this large-scale case, Scheme 4 could be solved effectively (we did not present it here). Further, 
the batch size at each worker can impact the convergence rate, but does not impact the final accuracy. 

\section{Conclusions}
In this paper, we consider a many-to-one wireless architecture for distributed learning at the network edge, where multiple edge devices collaboratively train a machine learning model, using local data, through a wireless channel. Observing  the unreliable nature of wireless connectivity, we design an integrated communication and learning scheme, where the local updates at edge devices are carefully crafted and compressed to match the wireless communication resources available. Specifically,  we propose SGD-based bandlimited coordinate descent algorithms  employing over-the-air computing,  in which a subset of k-coordinates of the gradient updates across edge devices are  selected by the receiver in each iteration and  then transmitted simultaneously over k sub-carriers. We analyze the convergence of the algorithms proposed, and characterize the effect of the communication error. Further, we  study joint optimization of  power allocation and learning rates therein to maximize the convergence rate. Our findings reveal that optimal power allocation across different sub-carriers should take into account both the gradient values and channel conditions. We then develop sub-optimal solutions amenable to implementation and verify our findings through numerical experiments.

\section*{Acknowledgements}
The authors thank Gautam Dasarathy for stimulating discussion in the early stage of this work. 
This work is supported in part by NSF Grants CNS-2003081, CNS-CNS-2003111, CPS-1739344 and ONR YIP N00014-19-1-2217.

\normalem
\bibliographystyle{IEEEtran}
\bibliography{library}

\begin{thebibliography}{10}
\providecommand{\url}[1]{#1}
\csname url@samestyle\endcsname
\providecommand{\newblock}{\relax}
\providecommand{\bibinfo}[2]{#2}
\providecommand{\BIBentrySTDinterwordspacing}{\spaceskip=0pt\relax}
\providecommand{\BIBentryALTinterwordstretchfactor}{4}
\providecommand{\BIBentryALTinterwordspacing}{\spaceskip=\fontdimen2\font plus
\BIBentryALTinterwordstretchfactor\fontdimen3\font minus
  \fontdimen4\font\relax}
\providecommand{\BIBforeignlanguage}[2]{{%
\expandafter\ifx\csname l@#1\endcsname\relax
\typeout{** WARNING: IEEEtran.bst: No hyphenation pattern has been}%
\typeout{** loaded for the language `#1'. Using the pattern for}%
\typeout{** the default language instead.}%
\else
\language=\csname l@#1\endcsname
\fi
#2}}
\providecommand{\BIBdecl}{\relax}
\BIBdecl

\bibitem{Zhu2018a}
G.~Zhu, D.~Liu, Y.~Du, C.~You, J.~Zhang, and K.~Huang, ``Towards an intelligent
  edge: Wireless communication meets machine learning,'' \emph{arXiv preprint
  arXiv:1809.00343}, 2018.

\bibitem{Goldenbaum2013}
M.~Goldenbaum and S.~Stanczak, ``Robust analog function computation via
  wireless multiple-access channels,'' \emph{IEEE Transactions on
  Communications}, vol.~61, no.~9, pp. 3863--3877, 2013.

\bibitem{Abari2016}
O.~Abari, H.~Rahul, and D.~Katabi, ``Over-the-air function computation in
  sensor networks,'' \emph{arXiv preprint arXiv:1612.02307}, 2016.

\bibitem{Alistarh2016}
D.~Alistarh, D.~Grubic, J.~Li, R.~Tomioka, and M.~Vojnovic, ``Qsgd:
  Communication-efficient sgd via gradient quantization and encoding,'' in
  \emph{Advances in Neural Information Processing Systems}, 2017, pp.
  1709--1720.

\bibitem{Wen2017}
W.~Wen, C.~Xu, F.~Yan, C.~Wu, Y.~Wang, Y.~Chen, and H.~Li, ``Terngrad: Ternary
  gradients to reduce communication in distributed deep learning,'' in
  \emph{Advances in Neural Information Processing Systems}, 2017, pp.
  1509--1519.

\bibitem{Bernstein2018a}
J.~Bernstein, Y.-X. Wang, K.~Azizzadenesheli, and A.~Anandkumar, ``Signsgd:
  Compressed optimisation for non-convex problems,'' in \emph{International
  Conference on Machine Learning}, 2018, pp. 559--568.

\bibitem{Wu2018}
J.~Wu, W.~Huang, J.~Huang, and T.~Zhang, ``Error compensated quantized sgd and
  its applications to large-scale distributed optimization,'' in
  \emph{International Conference on Machine Learning}, 2018, pp. 5321--5329.

\bibitem{Aji2017}
A.~F. Aji and K.~Heafield, ``Sparse communication for distributed gradient
  descent,'' in \emph{Proceedings of the 2017 Conference on Empirical Methods
  in Natural Language Processing}, 2017, pp. 440--445.

\bibitem{Stich2018}
S.~U. Stich, J.-B. Cordonnier, and M.~Jaggi, ``Sparsified sgd with memory,'' in
  \emph{Advances in Neural Information Processing Systems}, 2018, pp.
  4447--4458.

\bibitem{Alistarh2018}
D.~Alistarh, T.~Hoefler, M.~Johansson, N.~Konstantinov, S.~Khirirat, and
  C.~Renggli, ``The convergence of sparsified gradient methods,'' in
  \emph{Advances in Neural Information Processing Systems}, 2018, pp.
  5973--5983.

\bibitem{Konecny2016}
J.~Kone{\v{c}}n{\`y}, H.~B. McMahan, F.~X. Yu, P.~Richt{\'a}rik, A.~T. Suresh,
  and D.~Bacon, ``Federated learning: Strategies for improving communication
  efficiency,'' \emph{arXiv preprint arXiv:1610.05492}, 2016.

\bibitem{Stich2018a}
S.~U. Stich, ``Local sgd converges fast and communicates little,'' in
  \emph{ICLR 2019 International Conference on Learning Representations}, 2019.

\bibitem{Dong2018}
J.~Dong, Y.~Shi, and Z.~Ding, ``Blind over-the-air computation and data fusion
  via provable wirtinger flow,'' \emph{arXiv preprint arXiv:1811.04644}, 2018.

\bibitem{Liu2019}
W.~Liu and X.~Zang, ``Over-the-air computation systems: Optimization, analysis
  and scaling laws,'' \emph{arXiv preprint arXiv:1909.00329}, 2019.

\bibitem{Wen2018}
D.~Wen, G.~Zhu, and K.~Huang, ``Reduced-dimension design of {MIMO} over-the-air
  computing for data aggregation in clustered iot networks,'' \emph{IEEE
  Transactions on Wireless Communications}, vol.~18, no.~11, pp. 5255--5268,
  2019.

\bibitem{Zhu2018b}
G.~Zhu and K.~Huang, ``{MIMO} over-the-air computation for high-mobility
  multi-modal sensing,'' \emph{IEEE Internet of Things Journal}, 2018.

\bibitem{Cao2019}
X.~Cao, G.~Zhu, J.~Xu, and K.~Huang, ``Optimal power control for over-the-air
  computation in fading channels,'' \emph{arXiv preprint arXiv:1906.06858},
  2019.

\bibitem{Zhu2018}
G.~Zhu, Y.~Wang, and K.~Huang, ``Broadband analog aggregation for low-latency
  federated edge learning,'' \emph{IEEE Transactions on Wireless
  Communications}, 2019.

\bibitem{Yang2019}
K.~Yang, T.~Jiang, Y.~Shi, and Z.~Ding, ``Federated learning via over-the-air
  computation,'' \emph{arXiv preprint arXiv:1812.11750}, 2018.

\bibitem{Zeng2019}
Q.~Zeng, Y.~Du, K.~K. Leung, and K.~Huang, ``Energy-efficient radio resource
  allocation for federated edge learning,'' \emph{arXiv preprint
  arXiv:1907.06040}, 2019.

\bibitem{Amiri2019}
M.~M. Amiri and D.~G{\"u}nd{\"u}z, ``Machine learning at the wireless edge:
  Distributed stochastic gradient descent over-the-air,'' \emph{arXiv preprint
  arXiv:1901.00844}, 2019.

\bibitem{Amiri2019a}
------, ``Federated learning over wireless fading channels,'' \emph{arXiv
  preprint arXiv:1907.09769}, 2019.

\bibitem{Amiri2019c}
------, ``Over-the-air machine learning at the wireless edge,'' in \emph{2019
  IEEE 20th International Workshop on Signal Processing Advances in Wireless
  Communications (SPAWC)}.\hskip 1em plus 0.5em minus 0.4em\relax IEEE, 2019,
  pp. 1--5.

\bibitem{Ahn2019}
J.-H. Ahn, O.~Simeone, and J.~Kang, ``Wireless federated distillation for
  distributed edge learning with heterogeneous data,'' in \emph{2019 IEEE 30th
  Annual International Symposium on Personal, Indoor and Mobile Radio
  Communications (PIMRC)}.\hskip 1em plus 0.5em minus 0.4em\relax IEEE, 2019,
  pp. 1--6.

\bibitem{Sery2019}
T.~Sery and K.~Cohen, ``On analog gradient descent learning over multiple
  access fading channels,'' \emph{arXiv preprint arXiv:1908.07463}, 2019.

\bibitem{Karimireddy2019}
S.~P. Karimireddy, Q.~Rebjock, S.~U. Stich, and M.~Jaggi, ``Error feedback
  fixes signsgd and other gradient compression schemes,'' \emph{arXiv preprint
  arXiv:1901.09847}, 2019.

\bibitem{Ivkin2019}
N.~Ivkin, D.~Rothchild, E.~Ullah, V.~Braverman, I.~Stoica, and R.~Arora,
  ``Communication-efficient distributed sgd with sketching,'' \emph{arXiv
  preprint arXiv:1903.04488}, 2019.

\bibitem{shi2019distributed}
S.~Shi, Q.~Wang, K.~Zhao, Z.~Tang, Y.~Wang, X.~Huang, and X.~Chu, ``A
  distributed synchronous sgd algorithm with global top-k sparsification for
  low bandwidth networks,'' in \emph{2019 IEEE 39th International Conference on
  Distributed Computing Systems (ICDCS)}.\hskip 1em plus 0.5em minus
  0.4em\relax IEEE, 2019, pp. 2238--2247.

\bibitem{resnetpaper}
K.~He, X.~Zhang, S.~Ren, and J.~Sun, ``Deep residual learning for image
  recognition,'' \emph{arXiv preprint arXiv:1512.03385}, 2015.

\end{thebibliography}

\newpage
\onecolumn
\appendix

\section{Proofs}

\subsection{Proof of Theorem~\ref{thm:convergence}}
We here restate equations \eqref{eqn:updatesimplified} and \eqref{eqn:gcsimplified} as follows:
\begin{align}
    {w}_{t+1} =& w_t-[C_t(\gamma g_t(w_t)+r_t)+\epsilon_t]\\
    \hat{w}_{t+1}=& w_t-C_t(\gamma g_t(w_t)+r_t)
\end{align}
It is clear that \({w}_{t+1}=\hat{w}_{t+1}-\epsilon_t\). For convenience, we define \(\tilde{w}_t= {w}_{t} - r_t= \hat{w}_t-r_t- \epsilon_{t-1} \).
It can be shown that 
\(\tilde{w}_{t+1}=  \tilde{w}_{t} -  \gamma g_t(w_t) - \epsilon_{t} \).

\begin{align}
   \mathbb{E}_t [ f(\tilde{w}_{t+1}) ] 
   \leq& f(\tilde{w}_{t} )+<\nabla f(\tilde{w}_{t}),\mathbb{E}_t[\tilde{w}_{t+1}-\tilde{w}_{t}]>  +\frac{L}{2}\mathbb{E}_t[|| \tilde{w}_{t+1}-\tilde{w}_{t} ||^2]\\
    =& f(\tilde{w}_{t})- <\nabla f(\tilde{w}_{t}),\gamma \mathbb{E}_t[g_t (w_t)]  + \mathbb{E}_t[\epsilon_t] > + \frac{L}{2}\mathbb{E}_t[||   \gamma g_t(w_t) ||^2]\nonumber\\ &\hspace*{0.5in}+\frac{L}{2}\mathbb{E}_t[||\epsilon_t||^2] + L\mathbb{E}_t[< \gamma g_t(w_t) ,\epsilon_t>]\\
    =& f(\tilde{w}_{t}) -<\nabla f({w}_{t}),\gamma \mathbb{E}_t[g_t(w_t)] + \mathbb{E}_t[\epsilon_t]>  - <\nabla f(\tilde{w}_{t})-\nabla f({w}_{t}),  \gamma \mathbb{E}_t[g_t(w_t)] + \mathbb{E}_t[\epsilon_t]> \nonumber\\
    &\hspace*{0.5in} + \frac{L}{2}\mathbb{E}_t[||\epsilon_t||_2^2] + L\mathbb{E}_t[<\gamma g_t(w_t),\epsilon_t>] + \frac{L}{2}\mathbb{E}_t[||   \gamma g_t(w_t) ||^2 \\
    \leq& f(\tilde{w}_{t}) - \gamma \lVert \nabla f(w_t) \rVert_2^2 - \langle \nabla f(w_t), \mathbb{E}_t[\epsilon_t] \rangle + \frac{\rho}{2} \lVert \gamma \nabla f(w_t)+\mathbb{E}_t[\epsilon_t] \rVert_2^2+\frac{L^2}{2\rho}\mathbb{E}_t[\lVert r_t \rVert_2^2] \nonumber\\
    &\hspace*{0.5in}+ \frac{L}{2}\mathbb{E}_t[\lVert \epsilon_t \rVert_2^2] + L\langle \nabla f(w_t), \mathbb{E}_t[\epsilon_t] \rangle + \frac{L\gamma^2}{2} \mathbb{E}_t \lVert g_t(w_t) \rVert_2^2 \\
    \leq& f(\tilde{w}_{t}) - \gamma \lVert \nabla f(w_t) \rVert_2^2 + (L-1) \lVert \nabla f(w_t) \rVert\lVert \mathbb{E}_t[\epsilon_t] \rVert + \frac{\rho}{2}\left( \gamma^2\lVert \nabla f(w_t) \rVert_2^2 + \lVert \mathbb{E}_t[\epsilon_t] \rVert_2^2+2\gamma \langle \nabla f(w_t), \mathbb{E}_t[\epsilon_t] \rangle \right) \nonumber\\
    &\hspace*{0.5in}+ \frac{L^2}{2\rho} \mathbb{E}_t[\lVert r_t \rVert_2^2] + \frac{L}{2}\mathbb{E}_t[\lVert\epsilon_t\rVert_2^2] + \frac{L\gamma^2}{2} G^2 \\
    \leq& f(\tilde{w}_{t}) - \gamma \lVert \nabla f(w_t) \rVert_2^2 + (L-1+ 2 \gamma) \lVert \nabla f(w_t) \rVert\lVert \mathbb{E}_t[\epsilon_t] \rVert + \frac{\gamma^2 \rho }{2}\lVert \nabla f(w_t) \rVert_2^2\nonumber\\
    &\hspace*{0.5in}+ \frac{L^2}{2\rho} \mathbb{E}_t[\lVert r_t \rVert_2^2] + \lVert \mathbb{E}_t[\epsilon_t] \rVert_2^2  +\frac{L}{2}\mathbb{E}_t[\lVert\epsilon_t\rVert_2^2] + \frac{L\gamma^2}{2} G^2 \\
    =& f(\tilde{w}_{t}) -\gamma \left[ 1-\frac{\rho}{2}\gamma  \right] \lVert \nabla f(w_t) \rVert_2^2 + (L-1+2\gamma) \lVert \nabla f(w_t) \rVert \lVert \mathbb{E}_t[\epsilon_t] \rVert \nonumber\\
    &\hspace*{0.5in}+ \frac{L^2}{2\rho} \mathbb{E}_t[\lVert r_t \rVert_2^2] + \lVert \mathbb{E}_t[\epsilon_t] \rVert_2^2  +\frac{L}{2}\mathbb{E}_t[\lVert\epsilon_t\rVert_2^2] + \frac{L\gamma^2}{2} G^2
\end{align}
Based on \cite{Karimireddy2019}, we have that
\begin{align}
    \mathbb{E}_t[\lVert r_t \rVert_2^2]\leq \frac{4(1-\delta)}{\delta^2}\gamma^2 G^2.
\end{align}
It follows that 
\begin{align}
    & \frac{1}{T+1}\sum_{t=0}^T \left\{ \gamma(1-\frac{\rho}{2}\gamma) \lVert\nabla f(w_t)\rVert^2_2 - (L-1+2\gamma)\lVert \mathbb{E}_t[\epsilon_t] \rVert\lVert \nabla f(w_t) \rVert \right\}\nonumber\\
    &\hspace*{0.5in}\leq \frac{1}{T+1}[f(w_0)-f^*] + \frac{L^2}{\rho} \frac{2(1-\delta)}{\delta^2}\gamma^2 G^2  + \frac{L\gamma^2}{2} G^2 +  \frac{1}{T+1}\sum_{t=0}^T \left[  \lVert \mathbb{E}_t[\epsilon_t] \rVert_2^2  +\frac{L}{2}\mathbb{E}_t[\lVert\epsilon_t\rVert_2^2]  \right]
\end{align}

Through some further algebraic manipulation, we have that 
\begin{align}
    & \frac{1}{T+1}\sum_{t=0}^T \left(\lVert\nabla f(w_t)\rVert_2 - \frac{L-1+2\gamma}{\gamma (2-\rho\gamma)}\lVert \mathbb{E}_t[\epsilon_t] \rVert_2\right)^2
    \leq \frac{2}{T+1}\frac{f(w_0)-f^*}{\gamma(2-\rho\gamma)}+ \left(\frac{L}{\rho} \frac{2(1-\delta)}{\delta^2} + \frac{1}{2} \right) \frac{2L \gamma G^2}{ 2-\rho\gamma}  \nonumber\\
    &\hspace*{0.5in}+  \frac{1}{T+1}\sum_{t=0}^T \left[   \frac{L}{\gamma(2-\rho\gamma)}\mathbb{E}_t[\lVert\epsilon_t\rVert_2^2] + \left(1+ \frac{(L-1+2\gamma)^2}{\gamma^2 (2-\rho\gamma)^2} \right) \lVert \mathbb{E}_t[\epsilon_t] \rVert_2^2 \right] \label{contraction-result2}
\end{align}

For convenience,  let $\eta= \frac{L-1+2\gamma}{\gamma (2-\rho\gamma)} $, and define a contraction region as follows:
\[
C_{\beta} = \left\{ \lVert\nabla f(w_t)\rVert_2 \geq  (\eta + \Delta) \lVert \mathbb{E}_t[\epsilon_t] \rVert_2 \right\} .
\]
It follows from (\ref{contraction-result2}) that iterates in the BLCD algorithm returns to the contraction region infinitely often with probability one.
Further, when setting \(\gamma = \frac{1}{\sqrt{T+1}} \), we have that

  \begin{align}
    &   
     \frac{1}{T+1}\sum_{t=0}^T \left(\lVert\nabla f(w_t)\rVert_2 - \frac{L-1+2\gamma}{\gamma (2-\rho\gamma)}\lVert \mathbb{E}_t[\epsilon_t] \rVert_2\right)^2
    \leq \frac{f(w_0)-f^*}{(1-\frac{1}{2}\rho\gamma)}+  \frac{1}{\sqrt{T+1}} \left(\frac{L}{\rho} \frac{2(1-\delta)}{\delta^2} +  \frac{1}{2} \right) \frac{2L  G^2}{ 2-\rho\gamma}  \nonumber\\
    &\hspace*{0.5in}+ \frac{L}{(2-\rho\gamma)} \frac{1}{T+1}
    \sum_{t=0}^T    \mathbb{E}_t[\lVert\epsilon_t\rVert_2^2] + \left(1+ \frac{(L-1+2\gamma)^2}{\gamma^2 (2-\rho\gamma)^2} \right)
    \frac{1}{T+1}
    \sum_{t=0}^T   
     \lVert \mathbb{E}_t[\epsilon_t] \rVert_2^2 
\end{align}


\subsection{Solution of Problem (\textbf{P1-a})}
Since the problem (\textbf{P1-a}) is  convex, the Lagrangian function is given as:
	\begin{equation}
	L_{1a}(\bm{\alpha}, \bm{\lambda})=\text{MSE}_1(\bm{b}, \bm{\alpha}) +\sum_{k=1}^K\lambda_k \alpha_k.
	\end{equation}
Then,  the Karush-Kuhn-Tucker (KKT) conditions are given as follows:
	\begin{equation}
	\begin{split}
	\frac{\partial L_{1a}(\bm{\alpha}, \bm{\lambda})}{\partial \alpha_k} =& 2 \bigg( \sum_{m=1}^M b_{km} h_{km} x_{km}\bigg) \bigg(\sum_{m=1}^M \big( \alpha_k b_{km} h_{km} -\frac{1}{M} \big) x_{km}\bigg) + 2 \sigma^2\alpha_k + \lambda_k = 0,
	\end{split}
	\end{equation}
	\begin{equation}
	\lambda_k \geq 0, \quad \lambda_k \alpha^*_k=0, \quad \alpha^*_k \geq 0.
	\end{equation}
It follows that
	\begin{equation}
	\begin{split}
	\alpha^*_k &= \max\left\{\frac{\big(\sum_{m=1}^M x_{km}\big)\big(\sum_{m=1}^M b_{km} h_{km} x_{km} \big)}{M \big[\sigma^2 + \big(\sum_{m=1}^M b_{km} h_{km} x_{km} \big)^2\big]}, 0\right\} \triangleq \max\left\{\frac{ \bar{x}_k \beta_k }{\sigma^2 + \beta_k^2}\right\},
	\end{split}
	\end{equation}
	where the auxiliary variables are defined as $\beta_k= \sum_{m=1}^M b_{km} h_{km} x_{km}$ and $\bar{x}_k = \sum_{m=1}^M x_{km} / M$.

\subsection{Proof of Theorem \ref{thm:beta}}
\begin{proof}
Observing that the problem (\textbf{P2}) is defined only in terms of $b_k^2$, we define the auxiliary variables $\tilde{b}_k=b_k^2$, $\tilde{h}_k=h_k^2/ \sigma^2$ and $\tilde{x}_k=1/x_k^2$, and re-formulate (\textbf{P2}) as:
	\begin{equation*}
	\begin{aligned}
	\textbf{P2-1:\ } \min_{\tilde{\bm{b}}} \quad & \sum_{k=1}^K (\tilde{b}_k \tilde{h_k} + \tilde{x}_k)^{-1} \\
	\textrm{s.t.} \quad & \sum_{k=1}^{K} \frac{\tilde{b}_k}{\tilde{x}_k} \leq E \\
	& \tilde{b}_{k}\geq 0, \quad \forall k,
	\end{aligned}
	\end{equation*}
	which is convex and can be solved in closed form. Then, we have the Lagrangian  as
	\begin{equation}
	L_{22}(\tilde{\bm{b}}, \bm{\lambda}, \bm{\mu}) = \sum_{k=1}^K (\tilde{b}_k \tilde{h_k} + \tilde{x}_k)^{-1} + \lambda(\frac{\tilde{b}_k}{\tilde{x}_k}-E) - \sum_{k=1}^K \mu_k \tilde{b}_k,
	\end{equation}
	which leads to the following KKT conditions:
	\begin{equation*}
	\frac{\partial L_{12}(\tilde{\bm{b}}, \bm{\lambda}, \bm{\mu})}{\partial \tilde{b}_k} = - \tilde{h}_k (\tilde{b}_k \tilde{h}_k + \tilde{x}_k)^{-2} + \frac{\lambda}{\tilde{x}_k} - \mu_k = 0
	\end{equation*}
	\begin{equation*} \label{eq:powerconst}
	\sum_{k=1}^K \frac{\tilde{b}^*_k}{\tilde{x}_k} \leq E, \quad \tilde{b}^*_k \geq 0
	\end{equation*}
	\begin{equation*}
	\lambda \geq 0, \quad \mu_k \geq 0 
	\end{equation*}
	\begin{equation*}
	\lambda \bigg(E - \sum_{k=1}^K \frac{\tilde{b}^*_k}{\tilde{x}_k} \bigg)=0, \quad \mu_k \tilde{b}^*_k = 0.
	\end{equation*}
	For $\mu_k$=0, and $\lambda >0$, we have that
	\begin{equation} \label{eqn:singleusersolutionb}
	\tilde{b}_k^* = \max \bigg\{ \frac{\sqrt{\frac{\tilde{h}_k \tilde{x}_k}{\lambda}}-\tilde{x}_k}{\tilde{h}_k}, 0 \bigg\} = \bigg[ \frac{\sqrt{\frac{\tilde{h}_k \tilde{x}_k}{\lambda}}-\tilde{x}_k}{\tilde{h}_k} \bigg]^+
	\end{equation}
	with $E = \sum_{k=1}^K \frac{\tilde{b}^*_k}{\tilde{x}_k}$. By combining \eqref{eqn:singleusersolutionb} and $E = \sum_{k=1}^K \frac{\tilde{b}^*_k}{\tilde{x}_k}$, we obtain the following result:
	\begin{equation} \label{eqn:singleusersolutionlambda}
	E = \sum_{k=1}^K \bigg[ \frac{\sqrt{\tilde{h}_k \tilde{x}_k} \lambda'-\tilde{x}_k}{\tilde{h}_k \tilde{x}_k} \bigg]^+
	\end{equation}
	for $\lambda'=\sqrt{1/\lambda}$. \eqref{eqn:singleusersolutionlambda} can be solved by using the water-filling algorithm, where the solution can be found by increasing $\lambda '$ until the equality is satisfied. The optimal $\lambda'$ can be plugged int ${\tilde{b}^*_k}$ to yield $b^*_k=\sqrt{\tilde{b}^*_k}$ as a solution to (\textbf{P2}).  
\end{proof}

\subsection{Distributed Solutions towards Zero Bias and Variance Reduction (Scheme 2)}


The over-the-air gradient estimation requires a more comprehensive estimator design. To this end, a generalized optimization problem is defined for computing the optimal estimator. We define the MSE cost for the communication error, \(\epsilon_t\), in terms of the received signal \(\mathbf{y}=[y_1,y_2,\dots,y_K]\) as
\begin{align}
    \mathbb{E}[(\hat{{G}}-{G})^2] &= \mathbb{E}[(\boldsymbol{\alpha}\odot \mathbf{y}-\mathbf{G})^2] =\frac{1}{K}\sum_{k=1}^K \left[ \alpha_k\left( \sum_{m=1}^M b_{km}h_{km}x_{km}+n_k \right) - G_k \right]^2\label{q51}\\
    &= \frac{1}{K}\sum_{k=1}^K {\underbrace{\left( \mathbb{E}\left[\alpha_k\left( \sum_{m=1}^M b_{km}h_{km}x_{km}+n_k \right)\right] - \mathbb{E}\left[ \frac{1}{M}\sum_{m=1}^M x_{km} \right] \right)}_{e_k}}^2\nonumber\\
    &\hspace*{0.5in}+\underbrace{\mathbb{E}\left[ \left( \alpha_k\left( \sum_{m=1}^M b_{km}h_{km}x_{km}+n_k \right)-\mathbb{E}\left[\alpha_k\left( \sum_{m=1}^M b_{km}h_{km}x_{km}+n_k \right)\right] \right)^2 \right]}_{\nu_k^2},\label{q52}
\end{align}
where the estimator is \(\hat{G}=\alpha_k\left( \sum_{m=1}^M b_{km}h_{km}x_{km}+n_k \right)\) for \(k=1,2,\dots,K\). In \eqref{q52}, \(\alpha_k, b_{km}, h_{km}\) and \(n_k\) respectively denote the correction factor for recovering the \(k\)th dimension of the true gradient, the power allocation for the \(k\)th dimension of the local gradient \(x_{km}\) in the \(m\)th transmitter, the channel fading coefficient of the \(k\)th sub-channel between the \(m\)th transmitter and the receiver, and the thermal additive noise for the \(k\)th sub-channel. Further, \(\nu_k^2\) and \(e_k\) denote the estimator variance and bias, respectively. As apparent in \eqref{q52}, the minimization of the MSE cost does not ensure \(\hat{G}_k\) to be an unbiased estimator of \(G_k\). To resolve this issue, we formulate the unbiased optimization problem as
\begin{align}
    \underset{\{\alpha_k\},\{b_{km}\}}{\argmin}~ &\sum_{k=1}^K \nu_k^2(\alpha_k, \{b_{km}\})\label{q53}\\
    \text{subject to}~~ &e_k(\alpha_k, \{b_{km}\}) = 0, ~~\sum_{k=1}^K b_{km}^2x_{km}^2\leq E_m, ~~ b_{km}\geq 0, ~~\alpha_k\geq 0, &\forall k=1,\dots,K; ~~\forall m=1,\dots,M,\label{q54}
\end{align}
where \(E_m\) denotes the power budget of the \(m\)th transmitter.  We note that \(E_m\), \(\{x_{1m},x_{2m},\dots,x_{Km}\}\) and \(\{b_{1m},b_{2m},\dots,b_{Km}\}\) are only available to the \(m\)th transmitter  and \(\alpha_k, \forall k,\) are only available to the receiver.   The optimization problem defined in \eqref{q51}-\eqref{q52} can then be decomposed into two stages. In the first stage, each transmitter \(m\) utilizes all of its available power and the local gradient information to compute the optimal power allocation \(\{b_{1m},b_{2m},\dots,b_{Km}\}\). In the second stage, the receiver solves a consecutive optimization problem for finding the optimal \(\alpha_k\) for all \(k=1,\dots,K\). The optimization problem at each transmitter is formulated as
\begin{align}\label{prob2a}
    \underset{\{b_{km}\}_{k=1:K}}{\argmax} ~ &\zeta_m\\
    \text{subject to}~&\mathbb{E}\left[ \left(\zeta_mx_{km}-b_{km}h_{km}x_{km}\right)^2 \right] =0, ~~~\sum_{k=1}^K b_{km}^2x_{km}^2\leq E_m, ~~~b_{km}\geq 0, &\forall k=1,\dots,K.\label{q56}
\end{align}
The first constraint in \eqref{q56} ensures that there is no additive bias in the transmitted signal (i.e., the bias can be removed by a multiplicative factor), while the second constraint is the power constraint. The first constraint can be restated in a simpler form as \(\zeta_m = b_{km}h_{km}\), and then the solution can simply be obtained via the KKT conditions as
\begin{align}
    \text{Lagrangian:}\hspace{0.2in}& \mathcal{L}(\{b_{km}\}, \{\lambda_k\}, \vartheta, \{\beta_k\}) = \zeta_m - \sum_{k=1}^K \lambda_k (b_{km} h_{km}-\zeta_m) -\vartheta \left( \sum_{k=1}^K b_{km}^2x_{km}^2 - E_m \right) + \sum_{k=1}^K \beta_k b_{km}\nonumber\\
    \text{Stationarity:}\hspace{0.2in}& \frac{\partial\mathcal{L}}{\partial b_{km}} = 0 - \lambda_{k}h_{km} -2\vartheta x_{km}^2 b_{km} +\beta_k=0\rightarrow b_{km}=\frac{\beta_k-\lambda_k h_{km}}{2\vartheta x_{km}^2}\nonumber\\
    \text{Primal Feasibility:}\hspace{0.2in}&b_{km} h_{km}=\zeta_m, ~~\sum_{k=1}^K b_{km}^2x_{km}^2\leq E_m, ~~ b_{km}\geq 0, ~~\forall k=1\dots,K,\nonumber\\
    \text{Dual Feasibility:}\hspace{0.2in}& \vartheta\geq 0,~~\beta_k\geq 0, ~~\forall k=1,\dots,K,\nonumber\\
    \text{Comp. Slackness:}\hspace{0.2in}& \vartheta \left( \sum_{k=1}^K b_{km}^2x_{km}^2 - E_m \right) = 0,~~\sum_{k=1}^K \beta_k b_{km}=0.\nonumber
\end{align}
Then, the corresponding solution is given by 
\begin{align}\label{q57}
    \zeta_m^* = \sqrt{\frac{E_m}{\sum_{k=1}^K \frac{x_{km}^2}{h_{km}^2}}},&& b_{km}^*=\frac{\zeta_m}{h_{km}},~~~ \forall k=1,\dots,K.
\end{align}
\eqref{q57} illustrates that the \(m\)th transmitter utilizes all of its power budget to amplify its transmitted local gradient signal. Then, the corresponding received signal is equivalent to the \(\zeta_m^*\) times of the local gradient signal, inducing a multiplicative bias. Yet, this bias can be removed by multiplying the received signal with \(\boldsymbol{\alpha}\) in the receiver. However, the received signal at the receiver is a superposition of all transmitted signals because of the over-the-air transmission. Therefore, a single vector of optimal \(\boldsymbol{\alpha}\) must be computed for removing the bias and minimizing the estimator variance. To this end, in the second stage, the receiver solves the following optimization problem
\begin{align}
    \underset{\{\alpha_k\},}{\argmin}~ &\sum_{k=1}^K \nu_k^2(\alpha_k, \{b_{km}^*\})\label{q58}\\
    \text{subject to}~~ &e_k(\alpha_k, \{b_{km}^*\}) = 0, ~~\alpha_k\geq 0, &\forall k=1,\dots,K.\label{q59}
\end{align}
Next, we derive \(e_k(\alpha_k, \{b_{km}^*\})\) and \(\nu_k^2(\alpha_k, \{b_{km}^*\})\). For ease of exposition, the cumbersome steps are omitted here:
\begin{align}
    e_k(\alpha_k, \{b_{km}^*\}) &=\mathbb{E}\left[ \alpha_ky_k - \frac{1}{M}\sum_{m=1}^M x_{km} \right] = \mathbb{E}\left[ \alpha_k\left( \sum_{m=1}^M h_{km}b_{km}^*x_{km} + n_k \right) - \frac{1}{M}\sum_{m=1}^M x_{km} \right] \nonumber\\
    &=\alpha_k\left( \sum_{m=1}^M h_{km}b_{km}^*x_{km} \right) - \frac{1}{M}\sum_{m=1}^M x_{km} = \alpha_k\left( \sum_{m=1}^M \zeta_m^*x_{km} \right) - \frac{1}{M}\sum_{m=1}^M x_{km}\nonumber\\
    \nu_k^2(\alpha_k, \{b_{km}^*\}) &= \mathbb{E}\left[ \left( \alpha_k\left( \sum_{m=1}^M h_{km}b_{km}^*x_{km} +n_k \right) - \mathbb{E}\left[ \alpha_k\left( \sum_{m=1}^M h_{km}b_{km}^*x_{km} +n_k \right) \right] \right)^2 \right]\\
    &= \mathbb{E}\left[ \left( \alpha_k\left( \sum_{m=1}^M h_{km}b_{km}^*x_{km} - h_{km}b_{km}^*x_{km} +n_k \right)\right)^2 \right] = \alpha_k^2\sigma^2,
\end{align}
where the expectation is taken with respect to \(n_k\) for the realizations of all \(x_{km}\) and \(\sigma^2=\mathbb{E}[n_k^2]\). By solving the KKT conditions, we obtain the following solution
\begin{align}
    \alpha_k^*&= \frac{\frac{1}{M}\sum_{m=1}^M x_{km}}{\sum_{m=1}^M h_{km}b_{km}^*x_{km}} = \frac{\frac{1}{M}\sum_{m=1}^M x_{km}}{\sum_{m=1}^M \zeta_m^*x_{km}} .
\end{align}
From the implementation point of view, it is sensible to set $ \alpha_k^\dagger
\simeq \frac{1}{\sum_{m=1}^M \zeta_m^*}$ since $\{x_{km} \}$ is not available at the receiver. We herein also notice that \(\zeta_m^*\), for all \(m=1,\dots,M\), are not available at the receiver as well. Luckily, \(\alpha_k^\dagger\) is a function of \(\sum_{m=1}^M \zeta_m^*\), and hence a subchannel could be allocated for the over-the-air transmission of \(\sum_{m=1}^M \zeta_m^*\). Subsequently, it follows that the bias is given by
\begin{align}
    e_k(\alpha_k^*, \{b_{km}^*\}) &= \mathbb{E}\left[ \frac{1}{\sum_{m=1}^M \zeta_m^*}y_k - \frac{1}{M}\sum_{m=1}^M x_{km} \right] = \mathbb{E}\left[ \frac{1}{\sum_{m=1}^M \zeta_m^*}\left( \sum_{m=1}^M h_{km}b_{km}^*x_{km} + n_k \right) - \frac{1}{M}\sum_{m=1}^M x_{km} \right]\nonumber\\
    = \mathbb{E}&\left[ \frac{1}{\sum_{m=1}^M \zeta_m^*}\left( \sum_{m=1}^M \zeta_m^*x_{km} + n_k \right) - \frac{1}{M}\sum_{m=1}^M x_{km} \right] = \mathbb{E}\left[ \frac{\sum_{m=1}^M \zeta_m^*x_{km}+n_k}{\sum_{m=1}^M \zeta_m^*} - \frac{1}{M}\sum_{m=1}^M x_{km} \right]. \label{eq78}
    \end{align}
Assuming that the distributions of \(\{x_{km}\}\), for all \(k= 1,\dots,K ;m=1,\dots,M \), are identical across the subchannels and users, and so are  \(\{ h_{km} \}\), \(\zeta_m^*\) can be simplified. For ease of exposition, we denote \(\mathbb{E}[x_{km}^2] = \varphi^2+\bar{x}^2\) and \(\mathbb{E}\left[ \frac{1}{h_{km}^2} \right] = \varpi^2\). When the number of subchannels, \(K\), is large, we have that
\begin{align}
\zeta_m^* &= \frac{\sqrt{E_m}}{\sqrt{\sum_{k=1}^K \frac{x_{km}^2}{h_{km}^2}}} \underset{\text{when $K$ is large}}{\Longrightarrow} \zeta_m^* = \frac{\sqrt{E_m}}{\sqrt{K \mathbb{E}[x_{km}^2] \mathbb{E}\left[ \frac{1}{h_{km}^2} \right]}} = \frac{\sqrt{E_m}}{\sqrt{K (\varphi^2+\bar{x}^2) \varpi^2}} \\
e_k(\alpha_k^*, \{b_{km}^*\}) &= \mathbb{E}\left[ \frac{\sum_{m=1}^M \zeta_m^*x_{km}+n_k}{\sum_{m=1}^M \zeta_m^*} - \frac{1}{M}\sum_{m=1}^M x_{km} \right] = \sum_{m=1}^M \left[\frac{ \sqrt{E_m}}{\sum_{m=1}^M \sqrt{E_m}} - \frac{1}{M}\right] x_{km}. \label{eq82}
\end{align}
Then, the norm \(\lVert\mathbb{E}_t[ \epsilon_t ]\rVert_2^2\) in Theorem 1 can be expressed in terms of \(e_k, \forall k,\) as \(\lVert\mathbb{E}_t[ \epsilon_t ]\rVert_2^2=\sum_{k=1}^K\mathbb{E}_t[\epsilon_t]_k^2 = \sum_{k=1}^Ke_k^2\). 
Similarly, the variance \(\nu_k^2\) can also be computed as
\begin{align}
    \nu_k^2\left(\alpha_k^*,\{b_{km}^*\}\right) &= {\alpha_k^*}^2\sigma^2 = \left( \frac{1}{\sum_{m=1}^M \sqrt{\frac{E_m}{K\varpi(\varphi^2+\bar{x}^2)}}} \right)^2\sigma^2 = \frac{K\varpi^2(\varphi^2+\bar{x}^2)}{\left(\sum_{m=1}^M\sqrt{E_m}\right)^2}\sigma^2. \label{eq87}
\end{align}
Finally, the MSE cost in Theorem 1 can be written as \(\mathbb{E}_t[\lVert \epsilon_t \rVert_2^2]= \sum_{k=1}^K ({\nu}_k^2 + e_k^2)\).

\subsection{Alternative Formulations and Baselines}

{\bf A Receiver Centric Approach.} In what follows, we take a receiver-centric approach by selecting a fixed estimator, $\alpha_k=\frac{1}{Mp}, \forall k$, at the receiver. Given the fixed estimator, the MSE objective function is set as
\begin{equation}\label{qq60}
\begin{split}
\MSE(1/(Mp), \bm{b}) &=\sum_{k=1}^K \sum_{m=1}^M \big( \frac{b_{km} h_{km}}{Mp} -\frac{1}{M} \big)^2 x_{km}^2\nonumber\\
&+ \sum_{k=1}^K \sum_{m=1}^M \sum_{\substack{m'=1 \\ m' \neq m}}^M \big(  \frac{b_{km} h_{km}}{Mp} -\frac{1}{M} \big) \times \big(  \frac{b_{km'} h_{km'}}{Mp} -\frac{1}{M} \big) x_{km} x_{km'} +\sigma^2 \frac{K}{M^2 p^2}.
\end{split}
\end{equation}
We note that the first term can be decomposed for different devices and be solved in a distributed manner. The second term is coupled across different users, and the third term is only expressed in terms of the variable $p$. If there were no power constraints, the solution would be $b_{km} h_{km} = p$ for a very large $p$. With this intuition, it is sensible to assign most (if not all) of the multipliers $(\frac{ b_{km}h_{km}}{Mp} - \frac{1}{M})$  zero. That is to say,   $p$ should be selected so that it fits most users to provide enough power for the transmission of all the dimensions, while every user individually solves the power allocation problem using the first term for a given value of \(p\), i.e.,
\begin{equation*}
\begin{aligned}
\textbf{P2-1:\ } \min_{\{ b_{km} \}} \quad & \sum_{k=1}^K \big( \frac{b_{km} h_{km}}{Mp} -\frac{1}{M} \big)^2 x_{km}^2 \\
\textrm{s.t.} \quad & \sum_{k=1}^{K} b_{km}^2 x_{km}^2 \leq E.
\end{aligned}
\end{equation*}
    The optimal solution $\{ b_{km}^* \} $ to the problem (\textbf{P2-1}) can be obtained by finding $\lambda^*\geq 0$ satisfying the following two inequalities:
    \begin{equation}
    \sum_{k=1}^K (b_{km}^*)^2 x_{km}^2 \leq E, \quad b^*_{km}=\frac{x_{km}^2 p}{h_{km}(x_{km}^2 - M^2p^2 \lambda^*_m)}.
    \end{equation}
When \(\lambda^*=0\), then the solution is \(b_{km}^*=p/h_{km}\), the objective function is minimized to 0 and the power constraint is satisfied with strict inequality. On contrary, when \(\lambda^*>0\), then the power constraint is satisfied with equality and the optimal solution \(b_{km}^*\) deviates from \(p/h_{km}\). Then, the optimal solution $\{ b_{km}^* \} $ to the problem (\textbf{P2-1}) can be obtained by finding $\lambda^*> 0$ satisfying the following two equalities:
    \begin{equation}
    \sum_{k=1}^K (b_{km}^*)^2 x_{km}^2 = E, \quad b^*_{km}=\frac{x_{km}^2 p}{h_{km}(x_{km}^2 - M^2p^2 \lambda^*_m)}.
    \end{equation}


{\bf An equal power allocation approach.}
For comparison, we also consider the equal power allocation scheme, in which equal power is allocated to each dimension, i.e., we set $ b_{km}= b_m$. Enforcing the power constraint of the devices, e.g., \(\sum_{k=1}^{K} b_{km}^2 x_{km}^2 = E\), leads to the equal power solution that can be written as $b_m^* =  \sqrt{E/ {\sum_{k=1}^K x^2_{km}}}$.
Therefore, each device applies the power $b_m^*$ in each dimension, taking the advantage of the distribution of the data which is independent and identical. 

{\bf An Alternative Formulation to P3.}

When \(K\) is large, the channel coefficients \(\{h_{km}\}\) and the local gradients \(\{x_{km}\}\) are independent and identically distributed (i.i.d.) across all users and subchannels. Hence, the optimization problem \textbf{P3} can be further simplified via the law of large numbers as
\begin{align}\label{p4}
    \textbf{P4: } \underset{\tilde{b}}{\min}& ~\mathbb{E} \left[ \frac{1}{\tilde{b}_{km}\tilde{h}_{km}+\tilde{x}_{km}} \right] \nonumber\\
    \text{s.t. }& ~ \mathbb{E}[\tilde{b}_{km}/\tilde{x}_{km}]\leq \bar{E}_m,\nonumber\\
    &\tilde{b}_{km}\geq 0.
\end{align}
We here notice that the general law of large numbers only applies if all \(\tilde{b}_{km}\) are identically distributed. By further noting that \(\tilde{b}_{km}\) depends on \(\tilde{x}_{km}\) and \(\tilde{h}_{km}\), we conclude that the objective function for any \(k\) is identically distributed in \eqref{p4} . Then, \eqref{p4} simplifies as
\begin{align}
    \textbf{P4-1: } \underset{\tilde{b}_{km}}{\min}& ~\mathbb{E} \left[ \frac{1}{\tilde{b}_{km}\tilde{h}_{km}+\tilde{x}_{km}} \right] \nonumber\\
    \text{s.t. }& ~ \mathbb{E}\left[\frac{\tilde{b}_{km}}{\tilde{x}_{km}}\right]\leq \bar{E}_m,\nonumber\\
    &\tilde{b}_{km}\geq 0.
\end{align}
Subsequently, the KKT conditions are written as
\begin{align}
    \text{Lagrangian: } & \mathcal{L} = \mathbb{E} \left[ \frac{1}{\tilde{b}_{km}\tilde{h}_{km}+\tilde{x}_{km}} \right] +\lambda_m \left(\mathbb{E}\left[\frac{\tilde{b}_{km}}{\tilde{x}_{km}}\right]-\bar{E}_m\right)-\beta\tilde{b}_{km}\\
    \text{Stationarity: }& \frac{\partial \mathcal{L}}{\partial \tilde{b}_{km}} = \mathbb{E}\left[\frac{\lambda_m}{\tilde{x}_{km}} -\frac{\tilde{h}_{km}}{(\tilde{b}_{km}\tilde{h}_{km}+\tilde{x}_{km})^2} \right]=0\Leftarrow \frac{\lambda_m^*}{\tilde{x}_{km}} =\frac{\tilde{h}_{km}}{(\tilde{b}_{km}\tilde{h}_{km}+\tilde{x}_{km})^2}\\
    \text{Primal Feasibility: } & \mathbb{E}\left[\frac{\tilde{b}_{km}}{\tilde{x}_{km}}\right]\leq \bar{E}_m, ~~~ \tilde{b}_{km}\geq 0,\\
    \text{Dual Feasibility: } & \lambda_m\geq 0,~~~ \beta\geq 0,\\
    \text{Comp. Slackness: } & \lambda_m \left(\mathbb{E}\left[\frac{\tilde{b}_{km}}{\tilde{x}_{km}}\right]-\bar{E}_m\right)=0, ~~~ -\beta\tilde{b}_{km}=0.
\end{align}

Then, the solution is computed as follows
\begin{align}
    \tilde{b}_{km}=& \left[\sqrt{\frac{\tilde{x}_{km}}{\lambda_m^*\tilde{h}_{km}}}-\frac{\tilde{x}_{km}}{\tilde{h}_{km}}\right]^+\Rightarrow b_{km} = \sqrt{\left[ \frac{\sigma}{h_{km}|x_{km}|\sqrt{\lambda_m^*}}- \frac{\sigma^2}{x_{km}^2h_{km}^2} \right]^+}\\
    \lambda_m^*<&\frac{\tilde{h}_{km}}{\tilde{x}_{km}}\Rightarrow \tilde{b}_{km}>0 \text{ and } \lambda_m^*<\frac{h_{km}^2x_{km}^2}{\sigma^2}\Rightarrow b_{km}>0\label{q82}\\
    \mathbb{E}\left[\frac{\tilde{b}_{km}}{\tilde{x}_{km}}\right]=&\int_{0}^\infty\int_0^\infty \frac{\tilde{b}_{km}}{\tilde{x}_{km}} p(\tilde{h}_{km})p(\tilde{x}_{km})d\tilde{h}_{km}d\tilde{x}_{km}\\
    =&\int_{0}^\infty\int_0^{\lambda_m\tilde{x}_{km}} \left( \sqrt{\frac{1}{\lambda_m^*\tilde{h}_{km}\tilde{x}_{km}}}-\frac{1}{\tilde{h}_{km}} \right) p(\tilde{h}_{km})p(\tilde{x}_{km})d\tilde{h}_{km}d\tilde{x}_{km} =\bar{E}_m\\
    \mathbb{E}\left[{b}_{km}^2{x}_{km}^2\right]=& \int_{-\infty}^\infty \int_0^\infty \left[ \frac{\sigma}{h_{km}|x_{km}|\sqrt{\lambda_m^*}}- \frac{\sigma^2}{x_{km}^2h_{km}^2} \right]^+ x_{km}^2p(h_{km})p(x_{km})dh_{km}dx_{km}\\
    =& \int_{-\infty}^\infty \int_{\sqrt{\frac{\sigma^2\lambda_m^*}{x_{km}^2}}}^\infty \left( \frac{|x_{km}|\sigma}{h_{km}\sqrt{\lambda_m^*}}-\frac{\sigma^2}{h_{km}^2} \right) p(h_{km})p(x_{km})dh_{km}dx_{km}=\bar{E}_m.
\end{align}
By using the results above, the bias term, \(e_k\), is derived as
\begin{align}
    e_k(\alpha_k^*,\{b_{km}\}) =& \mathbb{E}\left[ \alpha_k^*y_k \right] - \frac{1}{M}\sum_{m=1}^M x_{km} = \mathbb{E}\left[ \sum_{m=1}^M \left( \alpha_k^*b_{km} h_{km} -\frac{1}{M} \right)x_{km} +\alpha_k^*n_k\right]\\
    =& \mathbb{E}\left[ \sum_{m=1}^M \left( \frac{1}{M}\sqrt{\left[ \frac{\sigma}{|x_{km}|h_{km}\sqrt{\lambda_m^*}}-\frac{\sigma^2}{x_{km}^2h_{km}^2} \right]^+}h_{km} -\frac{1}{M}\right)x_{km} +\frac{n_k}{M}  \right]\\
    =& \frac{1}{M}\sum_{m\in\mathcal{S}^\dagger}\left( \sqrt{\frac{\sigma h_{km}}{|x_{km}|\sqrt{\lambda_m^*}}-\frac{\sigma^2}{x_{km}^2}} -1 \right)x_{km},
\end{align}
where \(\mathcal{S}^\dagger\) represents the set of transmitters for which \eqref{q82} is satisfied. To compute the MSE cost, we compute the variance \({\nu}^2(\alpha_k^*,\{b_{km}\})\) by taking the expectation with respect to \(x_{km}\)
\begin{align}
    {\nu}_k^2(\alpha_k^*,\{b_{km}\}) =& \mathbb{E}\left[ \left( \sum_{m=1}^M \alpha_k^*b_{km}h_{km}x_{km}+\alpha_k^*n_k - \mathbb{E}\left[ \sum_{m=1}^M \alpha_k^*b_{km}h_{km}x_{km} \right] \right)^2 \right]\\
    &\hspace{-0.5in}= \mathbb{E}\left[ \left(\sum_{m=1}^M \frac{1}{M} \left\{ \sqrt{\left[ \frac{\sigma h_{km}}{\lvert x_{km} \rvert\sqrt{\lambda_m^*}}-\frac{\sigma^2}{x_{km}^2} \right]^+}x_{km}- \sqrt{\left[ \frac{\sigma h_{km}}{\lvert x_{km} \rvert\sqrt{\lambda_m^*}}-\frac{\sigma^2}{x_{km}^2} \right]^+}x_{km} \right\} + \frac{n_k}{M}\right)^2 \right]= \frac{\sigma^2}{M^2}.
\end{align}
Consequently, the cost \(\mathbb{E}_t\left[ \lVert \epsilon_t \rVert_2^2 \right]\) in Theorem 1 is computed as \(\mathbb{E}_t[\lVert \epsilon_t \rVert_2^2]= \sum_{k=1}^K (e_k^2 + {{\nu}}_k^2)\).

\end{document}